\documentclass[conference]{IEEEtran}
\IEEEoverridecommandlockouts
\usepackage{amsmath,amssymb,amsfonts}
\usepackage{algorithmic}
\usepackage{graphicx}
\usepackage{textcomp}
\usepackage{xcolor}

\usepackage[numbers,sort&compress,comma]{natbib}
\usepackage{amsmath,amsfonts}
\usepackage{algorithmic}
\usepackage{graphicx}
\usepackage{textcomp}
\usepackage{xcolor}
\usepackage{makecell}
\usepackage{mathrsfs}
\usepackage{amsthm}
\usepackage{epstopdf}
\usepackage{balance}
\usepackage{pdfx}

\usepackage{multirow}

\usepackage{epsfig,endnotes}
\usepackage{subfig}
\usepackage{grffile}
\usepackage[font=bf]{caption}
\usepackage{color, url}
\usepackage{xspace} 
\usepackage[ruled,linesnumbered]{algorithm2e}
\usepackage{epstopdf}
\usepackage{balance}
\usepackage{bm}
\usepackage{rotating}

\usepackage{enumitem}
\usepackage{makecell}

\usepackage{eso-pic}

\usepackage{bm}

\renewcommand{\mathbf}[1]{\bm{#1}}
\newcommand\CR[1]{\textcolor{black}{#1}}
\newcommand\CRR[1]{\textcolor{black}{#1}}

\usepackage{hyperref}

\newcommand{\myparatight}[1]{\smallskip\noindent{\bf {#1}:}~}

\allowdisplaybreaks

\def\BibTeX{{\rm B\kern-.05em{\sc i\kern-.025em b}\kern-.08em
    T\kern-.1667em\lower.7ex\hbox{E}\kern-.125emX}}
\begin{document}
\AddToShipoutPictureBG*{%
  \AtPageUpperLeft{%
    \setlength\unitlength{1in}%
    \hspace*{\dimexpr0.5\paperwidth\relax}
    \makebox(0,-0.75)[c]{In IEEE Symposium on Security and Privacy, 2022.}%
}}

\title{BadEncoder: Backdoor Attacks to Pre-trained Encoders in  Self-Supervised Learning}

\author{}
\author{ 
\IEEEauthorblockN{ Jinyuan Jia$^*$ \quad Yupei Liu$^*$ \quad Neil Zhenqiang Gong}
\IEEEauthorblockA{Duke University\\
\{jinyuan.jia, yupei.liu, neil.gong\}@duke.edu
\thanks{$^*$The first two authors made equal contribution.}
}
}

\maketitle

\begin{abstract}
Self-supervised learning  in computer vision aims to pre-train an image encoder 
using a large amount of unlabeled images or (image, text) pairs. The pre-trained image encoder can then be used as a feature extractor to build downstream classifiers for many downstream tasks with a small amount of or no labeled training data. In this work, we propose \emph{BadEncoder}, the first backdoor attack to self-supervised learning. In particular, our BadEncoder injects backdoors into a pre-trained image encoder such that the downstream classifiers built based on the backdoored image encoder for different downstream tasks simultaneously inherit the backdoor behavior.  
We formulate our BadEncoder as an optimization problem and we propose a gradient descent based method to solve it, which produces a backdoored image encoder from a clean one. Our extensive empirical evaluation results on multiple datasets show that our BadEncoder achieves high attack success rates while preserving the accuracy of the downstream classifiers. We also show the effectiveness of BadEncoder using two publicly available, real-world image encoders, i.e., Google's image encoder pre-trained on ImageNet  and OpenAI's  \emph{Contrastive Language-Image Pre-training (CLIP)} image encoder pre-trained on 400 million (image, text) pairs collected from the Internet. Moreover, we consider defenses including Neural Cleanse and MNTD (empirical defenses) as well as PatchGuard (a provable defense).  
Our results show that these defenses are insufficient to defend against BadEncoder, highlighting the needs for new defenses against our BadEncoder. Our code is publicly available at: \url{https://github.com/jjy1994/BadEncoder}.

\end{abstract}

\maketitle

\section{Introduction}
A key challenge of the conventional \emph{supervised learning} (or \emph{transfer learning}) is that they require a large amount of labeled training data for each classification task (or the teacher classification task). 
\emph{Self-supervised learning}~\cite{devlin2018bert,hadsell2006dimensionality,he2020momentum,chen2020simple,hjelm2018learning,grill2020bootstrap} is a new AI paradigm that aims to address the challenge. The self-supervised learning pipeline includes two key components, i.e., \emph{pre-training encoders} and \emph{building downstream classifiers}. 
For instance,  in the computer vision domain, the first component
 aims to pre-train an image encoder and (optionally) a text encoder using a large amount of \emph{unlabeled} images or (image, text) pairs (called \emph{pre-training dataset}). In the second component, the pre-trained image encoder is used as a feature extractor to build classifiers (called \emph{downstream classifiers}) for many downstream tasks with \emph{a small amount of} or \emph{no} labeled training data. 
 Self-supervised learning has achieved revolutionary and remarkable performance in various downstream tasks. For instance, OpenAI recently pre-trained CLIP~\cite{radford2021learning} on 400 million (image, text) pairs collected from the Internet, and without needing to use any labeled training data for the downstream  tasks, CLIP achieves accuracy that is competitive with the fully supervised classifiers for many downstream  tasks~\cite{radford2021learning}.

 However, existing studies~\cite{hadsell2006dimensionality,he2020momentum,chen2020simple,hjelm2018learning,grill2020bootstrap} on self-supervised learning mainly focus on designing new algorithms to pre-train encoders that achieve better performance for various downstream tasks, leaving the security of self-supervised learning in adversarial settings largely unexplored. 
We aim to bridge the gap in this work. 
In particular, we focus on backdoor attacks to self-supervised learning in the computer vision domain. An attacker aims to compromise the self-supervised learning pipeline such that \emph{backdoored downstream classifiers} are built for  attacker-chosen downstream classification tasks (called \emph{target downstream tasks}), and each backdoored downstream classifier predicts any input embedded with an attacker-chosen trigger as the corresponding attacker-chosen class (called \emph{target class}).

Existing backdoor attacks~\cite{gu2017badnets,chen2017targeted,liutrojaning2018,bagdasaryan2020blind} inject a backdoor into a classifier via compromising its training process, e.g., poisoning its labeled training data~\cite{gu2017badnets,chen2017targeted}, fine-tuning the classifier~\cite{liutrojaning2018}, or tampering the training algorithm~\cite{bagdasaryan2020blind}.  
In the context of self-supervised learning, these backdoor attacks require compromising the second component of the self-supervised learning pipeline, i.e., the training process of the downstream classifiers. Therefore, they are not applicable when a downstream classifier does not have a training process (i.e., the downstream task has no labeled training data) or its training process maintains integrity. 
Yao et al.~\cite{yao2019latent} proposed \emph{Latent Backdoor Attack (LBA)} to transfer learning. In particular, they inject a backdoor into a teacher classifier such that a student/downstream classifier built for a target downstream task is also backdoored. LBA can be extended to self-supervised learning via training a backdoored teacher classifier using the pre-trained image encoder and a large amount of labeled training data for both the target class and the non-target classes in the target downstream task. However, such a large amount of labeled training data may be unavailable. 
Moreover, as we will show in our experiments, even if such labeled training data is available, LBA achieves suboptimal attack effectiveness when extended to self-supervised learning.

\myparatight{Our work} In this work, we propose \emph{BadEncoder}, the first backdoor attack to self-supervised learning.   BadEncoder compromises the first component of the self-supervised learning pipeline, while assuming its  second component maintains integrity. Specifically,  BadEncoder  injects backdoors into a pre-trained image encoder such that the downstream classifiers built based on the backdoored encoder for the target downstream tasks simultaneously inherit the backdoor behavior. 

In particular, 
 our BadEncoder aims to achieve two goals: 1) a  backdoored downstream classifier  built based on the backdoored image encoder for a target downstream task should predict any input embedded with an attacker-chosen  trigger as the corresponding attacker-chosen target class, and 2) to make our BadEncoder stealthy, the backdoored downstream classifiers for both the target and non-target downstream tasks should maintain  accuracy for clean inputs. We call the two goals \emph{effectiveness goal} and \emph{utility goal}, respectively. An attacker can attack multiple target downstream tasks and multiple target classes for each target downstream task simultaneously. 
 We assume an attacker chooses a trigger for each (target downstream task, target class) pair; the attacker has access to one or more images (called \emph{reference inputs}) in the target class for each (target downstream task, target class) pair, e.g., the attacker can collect them from the Internet; and the attacker has access to a set of unlabeled images (called \emph{shadow dataset}), which may or may not be the pre-training dataset used to train the clean image encoder. 
 
We formulate our BadEncoder as an optimization problem. In particular, we propose an \emph{effectiveness loss} and an \emph{utility loss} to quantify the {effectiveness goal} and {utility goal}, respectively. Roughly speaking,  our effectiveness loss is smaller if the backdoored image encoder 1) produces more similar feature vectors for the reference inputs and the inputs in the shadow dataset embedded with the trigger for each (target downstream task, target class) pair, and 2) produces more similar feature vectors for the reference inputs with the clean image encoder. Our utility loss is smaller if the backdoored image encoder and the clean image encoder produce more similar feature vectors for each clean input in the shadow dataset. Our formulated optimization problem aims to craft a backdoored image encoder to minimize a weighted sum of the two loss terms. Moreover, we propose a gradient descent based method to solve the optimization problem, which produces a backdoored image encoder from a clean one.

We first evaluate our BadEncoder on CIFAR10, STL10, GTSRB, and SVHN datasets. In these experiments, we pre-train clean image encoders by ourselves and inject backdoors into them.  
Our experimental results show that BadEncoder can achieve high \emph{attack success rates}. 
For instance, the attack success rate is 99\% when the clean image encoder is pre-trained on CIFAR10 and the backdoored downstream classifier is built for GTSRB. Moreover,  
the accuracy loss (if any) of the backdoored downstream classifier incurred by BadEncoder is within 1\% in most cases. We also evaluate  BadEncoder on two publicly available, real-world image encoders. Specifically, we apply  BadEncoder to the image  encoder  pre-trained  on  ImageNet  and  released  by  Google~\cite{chen2020simple}  as  well  as  the   CLIP image  encoder pre-trained and released by OpenAI~\cite{radford2021learning}.  Our results indicate that  BadEncoder also achieves high attack success rates and maintains  accuracy of the downstream classifiers. 

We explore two state-of-the-art empirical defenses (i.e., Neural Cleanse~\cite{wang2019neural} and MNTD~\cite{xu2019detecting}) and a state-of-the-art provable defense (i.e., PatchGuard~\cite{xiang2020patchguard}) to mitigate our BadEncoder. In particular, both Neural Cleanse~\cite{wang2019neural} and MNTD~\cite{xu2019detecting} can detect whether a classifier is backdoored or not. Therefore, we apply them to detect  whether a downstream classifier is backdoored or not. Our experimental results indicate that they cannot detect the backdoored downstream classifiers. Our BadEncoder embeds a trigger/patch to an input to make the backdoored downstream classifier predict the target class. Therefore, we can use provable defenses against adversarial patches to mitigate our BadEncoder. In particular, we evaluate PatchGuard~\cite{xiang2020patchguard} which achieves the state-of-the-art certified accuracy against adversarial patches. Our experimental results indicate that PatchGuard provides insufficient robustness guarantees against our BadEncoder. Moreover, we extend MNTD to detect backdoored image encoders, and our results show that MNTD has low accuracy at detecting backdoored encoders. 

Our key contributions are summarized as follows:
\begin{itemize}
    \item We propose BadEncoder, the first backdoor attack to  self-supervised learning.
    
    \item We perform systematic evaluation for BadEncoder using multiple datasets. Moreover, we evaluate BadEncoder using two publicly available, real-world image encoders. 
    
    \item We explore three defenses to mitigate our BadEncoder. Our experimental results highlight that we need new defenses to defend against our BadEncoder. 
\end{itemize}

\section{Background on Self-supervised Learning}
\label{sec:background}
Self-supervised learning aims to pre-train an image encoder using a large amount of unlabeled data, and the pre-trained image encoder can then be used to build  classifiers for many downstream tasks (we consider image classification tasks in this work) with a small amount of or no labeled data. 
The image encoder can be pre-trained by a  resourceful service provider (e.g., Google, Facebook, OpenAI) and then shared with customers to build downstream classifiers. 
The self-supervised learning pipeline consists of two key components, i.e., \emph{pre-training an image encoder} and \emph{building a downstream classifier}. Next, we discuss the two components.

\subsection{Pre-training an Image Encoder} An image encoder is  a neural network which takes an image as input and outputs a \emph{feature vector} for it. Self-supervised learning pre-trains an image encoder using a large amount of unlabeled data which we call \emph{pre-training dataset}. The pre-training dataset could contain unlabeled images or  (image, text) pairs. Next, we discuss how to train image encoders based on unlabeled images or  (image, text) pairs.

\myparatight{Using unlabeled images} Among many methods~\cite{hadsell2006dimensionality,pathak2016context,noroozi2016unsupervised,he2020momentum,chen2020simple,hjelm2018learning,grill2020bootstrap} to pre-train an image encoder based on unlabeled images, \emph{contrastive learning}~\cite{hadsell2006dimensionality,he2020momentum,chen2020simple,hjelm2018learning,grill2020bootstrap} achieves state-of-the-art performance. Roughly speaking, the goal of contrastive learning is to learn an image encoder  that  produces similar feature vectors for different augmented versions of the same input image but produces dissimilar feature vectors for different input images. Specifically, contrastive learning quantifies such a goal using \emph{contrastive loss} defined on the unlabeled images.

We use SimCLR~\cite{chen2020simple}, a popular contrastive learning algorithm, as an example to illustrate the idea of contrastive learning. SimCLR contains several major components. The first component is \emph{data augmentation}, which contains a sequence of data augmentation operations such as random crop, random Gaussian blur, etc.. Given an input, this component produces an \emph{augmented input} via sequentially applying these operations to the input. The second component is an \emph{image encoder}, which outputs a feature vector for an input or an augmented input. The third component is a \emph{projection head}, which can be a multilayer perceptron (MLP) and maps a feature vector to a \emph{latent vector} that is used to define contrastive loss. 

Given a batch of $N$ input images, SimCLR creates $2\cdot N$ augmented inputs via applying data augmentation twice for each input. Two augmented inputs form a \emph{positive pair} if they were augmented from the same input, and they form a \emph{negative pair} otherwise.  
Roughly speaking, SimCLR learns the image encoder and projection head to maximize the cosine similarities between the latent vectors of the positive pairs and minimize those of the negative pairs. Formally, SimCLR defines contrastive loss  for a positive pair $(i,j)$  as follows: 
\begin{align}
    \ell_{i,j} = - \log(\frac{\exp(sim(\mathbf{z}_i, \mathbf{z}_j)/\tau)}{\sum_{k=1}^{2N} \mathbb{I}(k \neq i) \cdot \exp(sim(\mathbf{z}_i, \mathbf{z}_k)/\tau) } ),
\end{align}
where $\mathbf{z}_i$ and $\mathbf{z}_j$ are the latent vectors of the positive pair, $\mathbf{z}_i$ and $\mathbf{z}_k (k \neq i, j)$ are the latent vectors of a negative pair, $sim(\cdot,\cdot)$ measures the cosine similarity between two latent vectors, $\mathbb{I}$ is an indicator function, $\tau$ represents a temperature parameter, and $\exp$ is the natural exponential function.  The final contrastive loss is the sum of the contrastive loss for all positive pairs. SimCLR learns the image encoder and projection head via minimizing the final contrastive loss.

\myparatight{Using unlabeled (image, text) pairs} Another family of self-supervised learning methods~\cite{srivastava2012multimodal,joulin2016learning,thomee2016yfcc100m,li2020unicoder,radford2021learning} try to pre-train an image encoder based on unlabeled (image, text) pairs. We note that, other than an image encoder, these methods also pre-train a text encoder, which takes a text (e.g., a sentence) as an input and outputs a feature vector for it.  
Recently, Radford et al.~\cite{radford2021learning} proposed \emph{Contrastive Language-Image Pre-training (CLIP)} which learns an image encoder and a text encoder using 400 million (image, text) pairs collected from the Internet, and achieves state-of-the-art performance. Given a batch of $N$ (image, text) pairs (we call them \emph{positive pairs}), CLIP constructs $N\times (N -1)$  negative (image, text) pairs, each of which consists of the image in one positive pair and the text in another positive pair. CLIP calculates a cosine similarity for each positive/negative (image, text) pair. Specifically, given a (image, text) pair, the image encoder produces a feature vector for the image and the text encoder produces a feature vector for the text; and CLIP calculates the cosine similarity between the two feature vectors for the pair. Roughly speaking, CLIP learns the image encoder and text encoder to maximize the cosine similarity for the $N$ positive pairs and minimize the cosine similarity for the $N\times (N -1)$  negative pairs.

\subsection{Building a Downstream Classifier}
The pre-trained image encoder can be used as a feature extractor to build classifiers (called \emph{downstream classifiers}) for many downstream tasks. 
Depending on whether a labeled training dataset is required or not, a downstream classifier could be \emph{multi-shot classifier} or \emph{zero-shot classifier}. A multi-shot classifier is trained via supervised learning with multiple labeled training examples, while a zero-shot classifier requires zero labeled training examples. Note that zero-shot classifier requires both  an image encoder and a text encoder, i.e., zero-shot classifier is only applicable when the image/text encoder is pre-trained based on (image, text) pairs.  

\myparatight{Multi-shot classifier}
In this scenario, we have multiple labeled training examples (we call them \emph{downstream dataset}) for the downstream task. 
We use the pre-trained image encoder to produce a feature vector for each image in the downstream dataset, and then we  train a classifier via the standard supervised learning. 
Given a testing input, we  first use the pre-trained image encoder to produce a feature vector for it and then use the  trained classifier to predict a label for it.

\myparatight{Zero-shot classifier} In this scenario, we have zero labeled training examples for the downstream task, but both an image encoder and a text encoder are available. To build a zero-shot classifier for a downstream task, we first construct a \emph{context sentence} for each class of the downstream task. For instance, Radford et al.~\cite{radford2021learning} showed that a context sentence like ``A photo of a \{class name\}'' can be a good default template that outperforms the baseline of using only label text (i.e., ``\{class name\}''), where ``\{class name\}'' can be ``stop sign'', ``yield'', ``speed limit'', etc. when the downstream task is traffic sign classification. Moreover, they also found that the accuracy of a zero-shot classifier can be further improved by customizing the context sentence for each downstream task. We can construct $c$ context sentences for the $c$ classes of the downstream task. We use the text encoder to produce a feature vector for each context sentence. Given a testing image, we use the image encoder to produce a feature vector for it. Then, the zero-shot classifier predicts the testing image as the class whose context sentence's feature vector has the largest cosine similarity with the image's feature vector.

\section{Threat Model}






We characterize our threat model with respect to attacker's goals, background knowledge, and capabilities. 

\myparatight{Attacker's goals} We consider an attacker aims to inject backdoors into a pre-trained image encoder such that a downstream classifier built based on the backdoored image encoder makes attacker-chosen predictions for 
inputs embedded with an attacker-chosen trigger. 
We consider attacking image encoder instead of text encoder so our attacks are applicable to both images and (image, text)  pairs based self-supervised learning. 
In particular, the attacker first selects some downstream tasks that he/she aims to target. We call such downstream tasks  \emph{target downstream tasks}. For each target downstream task, an attacker could target one or more of its classes, which we call \emph{target classes}. For instance, when the target downstream task is traffic sign recognition, the target classes could be  ``priority traffic sign'', ``stop sign'', etc.. For simplicity, we use $(T_i, y_i)$ to denote a (target downstream task, target class) pair, where $i=1, 2, \cdots, t$. Note that multiple $T_i$'s may represent the same target downstream task when the attacker selects multiple target classes for it.   For each $(T_i, y_i)$ pair, the attacker  selects a backdoor trigger  $\mathbf{e}_i$, e.g., a patch located at the bottom right corner of an input image. 

The attacker aims to craft a backdoored image encoder based on a clean image encoder to achieve two goals, i.e., \emph{effectiveness goal} and \emph{utility goal}. 
\begin{itemize}
    \item {\bf Effectiveness goal.} The effectiveness goal means that, when a downstream classifier (called \emph{backdoored downstream classifier}) is built based on the backdoored image encoder for  a target downstream task $T_i$,  the backdoored downstream classifier should predict the target class $y_i$ for any input embedded with the trigger $\mathbf{e}_i$.  The backdoored downstream classifiers  for the target downstream tasks should simultaneously have such backdoor behavior. 
    \item {\bf Utility goal.} The utility goal means that, the backdoored image encoder should maintain utility to be stealthy. In particular, for any target or non-target downstream task, a downstream classifier built based on the backdoored image encoder should be as accurate as a downstream classifier built based on the clean image encoder for clean testing inputs. 
\end{itemize}

\myparatight{Attacker's background knowledge and capabilities}  We consider two possible attackers: 1) an untrusted service provider who  injects a backdoor into its pre-trained image encoder and shares the backdoored  encoder with downstream customers (e.g., makes the backdoored  encoder publicly available), and 2) a malicious third-party  who obtains a clean pre-trained image encoder from a service provider, injects backdoors into it, and shares the backdoored encoder with downstream customers (e.g., via republishing it for public download  on GitHub). Therefore, an attacker has access to a clean pre-trained image encoder.  We note that our attack is only applicable when a downstream customer uses an image encoder from an untrusted source and that our work shows one example of how this encoder could have been compromised.

Moreover, we assume an attacker has access to a set of unlabeled images, which we call \emph{shadow dataset}.  
In particular, we consider two scenarios depending on who is the attacker. In the first scenario,  the attacker is an untrusted service provider who pre-trains the image encoder and thus the attacker has access to the pre-training dataset, in which the attacker can use the pre-training dataset  as the shadow dataset.  In the second scenario where the attacker is a malicious third-party, the attacker may not have access to the pre-training dataset and thus the shadow dataset may not be the pre-training dataset. In this scenario, we will consider  shadow dataset that has or does not have the same distribution as the pre-training dataset. We also assume the attacker has access to some images (called \emph{reference inputs}) for each (target downstream task, target class) pair, e.g., the attacker can collect such images from the Internet. For instance, for the (traffic sign recognition, ``stop sign'') pair, the attacker can collect one or more stop sign images from the Internet as the reference inputs. Note that the reference inputs are not in the downstream dataset used to build downstream classifiers.  As we will see in Section~\ref{sec:attacks}, our BadEncoder uses the shadow dataset and reference inputs to inject backdoors into the image encoder.

We assume the attacker does not have access to the downstream dataset used to build downstream classifiers and cannot tamper the training process of the downstream classifiers.

\section{Design of BadEncoder}
\label{sec:attacks}

\begin{figure*}[!t]
	 \centering
{\includegraphics[width=0.9\textwidth]{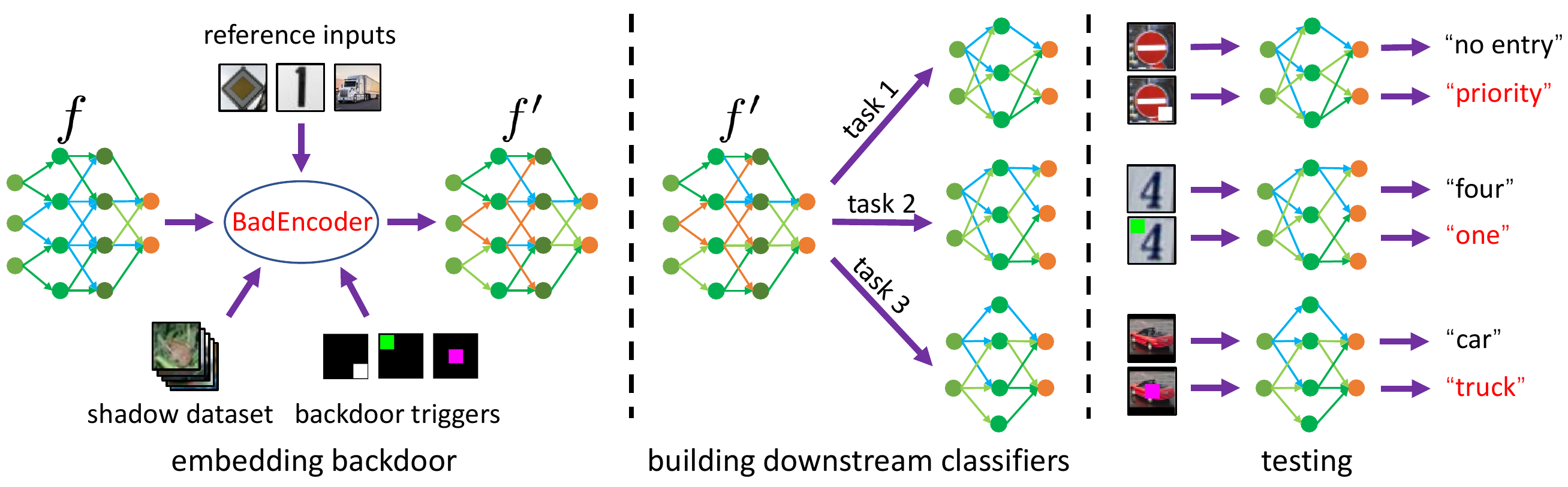}}
\caption{Overview of BadEncoder.}
\label{overview_of_our_attacks}
 \vspace{-6mm}
\end{figure*}

\subsection{Overview}
Figure~\ref{overview_of_our_attacks} shows an overview of BadEncoder. 
We aim to craft a backdoored image encoder from a clean one to achieve the effectiveness goal and utility goal. 
To achieve the effectiveness goal, our idea is to modify the clean image encoder such that 1) it produces similar feature vectors for the reference inputs and inputs in the attacker's shadow dataset embedded with the trigger for each (target downstream task, target class) pair, and 2) it produces similar feature vectors for the reference inputs with the clean image encoder. Therefore, a downstream classifier built based on our backdoored image encoder still predicts a reference input as the target class, and thus likely predicts any input embedded with the corresponding trigger as the target class. To achieve the utility goal, we modify the clean image encoder such that our backdoored image encoder and the clean image encoder produce similar feature vectors for each clean input in the shadow dataset. Therefore, a downstream classifier built based on our backdoored image encoder will maintain its accuracy for clean testing inputs. 

Formally, we formulate our BadEncoder as an optimization problem, solving which gives us a backdoored image encoder that achieves the two goals. In particular, we propose an \emph{effectiveness loss} and an \emph{utility loss} to quantify the two goals, respectively. Our optimization problem aims to minimize a weighted sum of the losses.

\subsection{Formulating our BadEncoder as an Optimization Problem}

We denote a clean pre-trained image encoder and our backdoored one as $f$ and $f'$, respectively. For each (target downstream task, target class) pair $(T_i, y_i)$, the attacker collects a set of reference inputs $\mathcal{R}_i=\{\mathbf{x}_{i1}, \mathbf{x}_{i2}, \cdots, \mathbf{x}_{ir_i}\}$ from the target class $y_i$, where $r_i$ is the number of reference inputs for the pair $(T_i, y_i)$ and $i=1,2,\cdots,t$. $\mathbf{e}_i$ is the attacker-chosen trigger for the  pair $(T_i, y_i)$. $\mathbf{x} \oplus \mathbf{e}_i$ means embedding the trigger $\mathbf{e}_i$ to an input $\mathbf{x}$. We call $\mathbf{x} \oplus \mathbf{e}_i$ a \emph{backdoored input}. We propose an effectiveness loss and an utility loss to quantify the effectiveness goal and utility goal, respectively. Next, we discuss them. 

\myparatight{Effectiveness loss} To achieve the effectiveness goal, we craft a backdoored image encoder such that it produces similar feature vectors for the reference inputs $\mathcal{R}_i$ of a pair $(T_i, y_i)$ and any input in the shadow dataset $\mathcal{D}_s$ embedded with the trigger $\mathbf{e}_i$. Therefore, a backdoored downstream classifier built based on our backdoored image encoder is likely to predict the same label for the reference inputs $\mathcal{R}_i$ and any input embedded with the trigger $\mathbf{e}_i$. However, this alone does not guarantee that the backdoored downstream classifier predicts the target class $y_i$ for an input  embedded with the trigger $\mathbf{e}_i$, because  it may not correctly predict the target class for the reference inputs. Therefore, we further require that the backdoored image encoder produces similar  feature vectors for the reference inputs with the clean image encoder. 
Formally, our effectiveness loss consists of the following two terms:
\begin{align}
    L_0 &=-\frac{\sum_{i=1}^{t} \sum_{j=1}^{r_i}\sum_{\mathbf{x} \in \mathcal{D}_s} s(f'(\mathbf{x} \oplus \mathbf{e}_i), f'(\mathbf{x}_{ij}))}{ |\mathcal{D}_s|\cdot \sum_{i=1}^t r_{i}}, \\
    \label{equation_of_l1}
    L_1 &=  -\frac{\sum_{i=1}^{t} \sum_{j=1}^{r_i} s(f'(\mathbf{x}_{ij}), f(\mathbf{x}_{ij}))}{\sum_{i=1}^t r_{i}},
\end{align}
where $s(\cdot, \cdot)$ measures the similarity (e.g., cosine similarity) between two feature vectors, $|\mathcal{D}_s|$ represents the number of inputs in the shadow dataset, and the denominators in $L_0$ and $L_1$ are used to normalize the losses. Our effectiveness loss is a weighted sum of the two terms, i.e., $L_0 + \lambda_1 L_1$, where $\lambda_1$ is a hyperparameter to balance the two terms.  
$L_0$ is smaller if the backdoored image encoder $f'$ produces more similar feature vectors for the reference inputs and backdoored inputs in the shadow dataset, while $L_1$ is smaller if the backdoored image encoder and clean image encoder produce more similar feature vectors for the reference inputs.

\myparatight{Utility loss} Our BadEncoder aims to maintain the utility of the backdoored image encoder,  
i.e., maintain the accuracy of the downstream classifiers built based on our backdoored image encoder for clean inputs. 
 When both an image encoder and a text encoder are pre-trained using (image, text) pairs, a downstream classifier can be a zero-shot classifier. We note that our attack only embeds backdoor to an image encoder to be more generally applicable. Therefore, a backdoored zero-shot classifier and its clean version may predict different labels for a clean input if the backdoored image encoder and the clean image encoder produce different feature vectors for it. This is because the feature vectors produced by the text encoder do not change in the zero-shot classifiers. 
 Based on this observation, we require our backdoored image encoder and the clean image encoder to produce similar feature vectors for a clean input, e.g., a clean input in the shadow dataset. Specifically, our utility loss is smaller if our backdoored image encoder and the clean image encoder produce more similar feature vectors for a clean input in the shadow dataset. 
Formally, we define our utility loss as follows:
\begin{align}
    L_2 = - \frac{1}{|\mathcal{D}_s|} \cdot \sum_{\mathbf{x} \in \mathcal{D}_s} s(f'(\mathbf{x}), f(\mathbf{x})). 
\end{align}

\myparatight{Optimization problem} After defining the three loss terms $L_0$, $L_1$, and $L_2$, we formulate our BadEncoder as  an optimization problem. Specifically, our backdoored image encoder is a solution to the following optimization problem:
\begin{align}
\label{final_loss_term}
\min_{f'} L = L_0 + \lambda_1 \cdot L_1 + \lambda_2 \cdot L_2,
\end{align}
where $\lambda_1$ and $\lambda_2$ are two hyperparameters to balance these three loss terms. 
We will explore their impact on our BadEncoder in our evaluation. As our experimental results show, each of the three loss terms is necessary for our BadEncoder to achieve both the effectiveness goal and utility goal. Note that we can also jointly optimize the backdoored image encoder $f'$ and the backdoor triggers  $\mathbf{e}_i$'s (both locations of the triggers and their pixel values) in (\ref{final_loss_term}). However, we find that our BadEncoder with simple, physically realizable triggers (e.g., a white square located at the bottom right corner of an image) already achieves the two goals. Therefore, for simplicity, we don't optimize the  triggers in this work and leave such joint optimization as a future work.

\subsection{Solving the Optimization Problem}
An algorithm to solve the optimization problem in (\ref{final_loss_term}) is an attack to craft a backdoored image encoder. Our BadEncoder solves the optimization problem using gradient descent. 
Specifically, we initialize the backdoored image encoder as the clean image encoder. In each epoch, we sample a mini-batch of the shadow dataset, calculate the gradient of the loss $L$, and move the backdoored image encoder a small step (called learning rate) towards the inverse of the gradient. We repeat the process for $max\_epoch$ epochs. Algorithm~\ref{alg:example}  in Appendix shows our BadEncoder to solve the optimization problem, where the function \textsc{MiniBatch} samples a mini-batch of $bs$ inputs from the shadow dataset $\mathcal{D}_s$.

\section{Evaluation}
\label{sec:exp}
\subsection{Experimental Setup}

\subsubsection{Datasets} We use the following five image datasets.
\begin{itemize}
    
\item {\bf CIFAR10~\cite{krizhevsky2009learning}:} This dataset contains 50,000  training images and 10,000  testing images. Each image has a size of 32$\times$32$\times$3 and belongs to one of 10 classes. 

\item {\bf STL10~\cite{coates2011analysis}:} 
This dataset contains 5,000 labeled training images and 8,000 labeled testing images, each of which has a size of 96$\times$96$\times$3. Moreover, the dataset contains 10 classes and each image belongs to one of them. Besides the labeled training and testing images, the dataset also contains 100,000 unlabeled images.  

\item {\bf GTSRB~\cite{stallkamp2012man}:} This dataset contains 51,800 traffic sign images in 43 categories. Each image has a size of 32$\times$32$\times$3. The dataset is divided into 39,200 training images and 12,600 testing images. 

\item {\bf SVHN~\cite{netzer2011reading}:} In this dataset,  each image represents a digit from the house numbers in Google Street View. The size of each image is 32$\times$32$\times$3. Moreover, each image belongs to one of the 10 digits. This dataset has 73,257 training images and 26,032 testing images. According to~\cite{netzer2011reading}, they introduced some distracting digits to the sides of the digit of interest, i.e., SVHN is a noisy dataset. 

\item {\bf Food101~\cite{bossard14}:} This dataset contains   101,000 images of 101 food categories. We note that this dataset is only used to study the impact of the shadow dataset on BadEncoder. 
\end{itemize}

\subsubsection{Pre-training image encoders}\label{pre_training_image_encoder} When a dataset is used to pre-train an image encoder, we call it \emph{pre-training dataset}. In our experiments, we use CIFAR10 or STL10 as a pre-training dataset since they contain more images and are not noisy datasets. In particular, when CIFAR10  is a pre-training dataset, we use its training images (excluding the labels) to pre-train an image encoder; and  when STL10 is a pre-training dataset, we further consider its unlabeled images when pre-training an image encoder. Note that we do not use the testing images  when pre-training an image encoder because we reserve them to evaluate our BadEncoder when our shadow dataset is different from the pre-training dataset but they have the same distribution. Given a pre-training dataset, we use SimCLR~\cite{chen2020simple} to train a ResNet18~\cite{he2016deep} as an image encoder.  \CRR{Our implementation is based on the publicly available code  of SimCLR~\cite{simclr_url,simclr_url_pytorch}.} We train an image encoder for 1,000 epochs with the Adam optimizer and initial learning rate 0.001. 

\subsubsection{Training downstream classifiers} Given an (backdoored) image encoder pre-trained using a pre-training dataset, we use it to train downstream classifiers (i.e., multi-shot classifiers) for the remaining three datasets. When a dataset is used to train a downstream classifier, we call it \emph{downstream dataset}. For instance, given an image encoder pre-trained on CIFAR10, we use it to train downstream classifiers for downstream datasets STL10, GTSRB, and SVHN. In particular, we use a fully connected neural network with two hidden layers as a downstream classifier for a downstream dataset. The number of neurons in the two hidden layers are 512 and 256, respectively. When a dataset is treated as a downstream dataset, we use its training dataset to train a downstream classifier and we use its testing dataset to evaluate the downstream classifier. 
Specifically, we adopt the cross-entropy loss function and Adam optimizer when training a downstream classifier. Moreover, we train a downstream classifier for 500 epochs with an initial learning rate $0.0001$. When a downstream classifier is trained based on a backdoored image encoder, we call it \emph{backdoored downstream classifier}.

\subsubsection{Evaluation metrics} We use \emph{Clean Accuracy (CA)}, \emph{ Backdoored Accuracy (BA)}, and \emph{Attack Success Rate (ASR)} to evaluate our BadEncoder. CA and  BA measure the classification accuracies of a clean downstream classifier and a backdoored downstream classifier for the clean downstream testing dataset, respectively. 
ASR measures the fraction of trigger-embedded testing inputs that are predicted as the target class by a backdoored downstream classifier. Formally, we define them as follows:

\begin{itemize}

    \item {\bf Clean Accuracy (CA):} The CA of a clean downstream classifier is its classification accuracy for the clean testing images of the corresponding downstream dataset.

    \item {\bf  Backdoored Accuracy (BA):}  The BA of a backdoored downstream classifier is its classification accuracy for the clean testing images of the corresponding downstream dataset. 
    When the BA of a backdoored downstream classifier is similar to the CA of the corresponding clean downstream classifier, our attack preserves accuracy for the corresponding downstream task.

    \item {\bf Attack Success Rate (ASR):} Given a target downstream task and target class, we embed the attacker-chosen trigger to the testing images of the corresponding downstream dataset. The ASR of a backdoored downstream classifier is the fraction of such trigger-embedded testing images that are predicted as the target class by the backdoored downstream classifier. As a baseline,  we also consider \emph{Attack Success Rate-Baseline (ASR-B)}, which is the fraction of such trigger-embedded testing images that are predicted as the target class by the corresponding clean downstream classifier. ASR-B represents the attack success rate when the attacker does not inject backdoor to the pre-trained image encoder. 
    
\end{itemize}

\subsubsection{Parameter setting} 
\label{parameter_setting}
By default, we consider the attacker selects a single target downstream task/dataset and a single target class. 
We will also evaluate our BadEncoder when the attacker selects multiple target downstream tasks and/or target classes. Moreover, we select ``airplane'', ``truck'', ``priority sign'', and ``digit one'' as the target class for the four datasets CIFAR10, STL10, GTSRB, and SVHN, respectively. Unless otherwise mentioned, we adopt CIFAR10 as the default pre-training dataset and STL10 as the default target downstream dataset. Note that we resize each image in STL10 dataset to $32 \times 32 \times 3$ to be consistent with other datasets.

Our BadEncoder has the following parameters: $\lambda_1$, $\lambda_2$, shadow dataset, trigger, and reference inputs. 
Unless otherwise mentioned, we use the following default parameter settings: $\lambda_1 = 1$ and $\lambda_2 = 1$; shadow dataset includes $50,000$ images sampled from the pre-training dataset; and following previous work on backdoor attacks to classifiers~\cite{wang2019neural}, we use a $10 \times 10$ white square located at the bottom right corner of an input image as the trigger. The size (e.g., $10 \times 10$) of a trigger refers to its height and width. 
By default, we assume one reference input for a (target downstream task, target class) pair. We collected the reference input for each target downstream dataset with the default target class from the Internet. Figure~\ref{four_attack_inputs} in Appendix shows our default reference input for each downstream dataset. We consider the collected reference input is correctly classified as the target class by the backdoored downstream classifier.   An attacker can select multiple reference inputs from the target class, and 
we will show our BadEncoder is effective once at least one reference input is correctly classified by the backdoored downstream classifier. In experiments, we use a randomly augmented version of $x_{ij}$ in $f'(x_{ij})$ in Equation~\ref{equation_of_l1} in each mini-batch. \CRR{Moreover, we freeze the parameters in the batch normalization layers of an encoder when embedding backdoor into it.}

We will explore the impact of each parameter on BadEncoder. 
We adopt cosine similarity to measure the similarity of two inputs' feature vectors outputted by an image encoder  in our loss terms. We fine-tune a pre-trained image encoder for 200 epochs using Algorithm~\ref{alg:example} with learning rate $0.001$ and batch size $256$ to inject the backdoor. 

\begin{table}[tp]\renewcommand{\arraystretch}{1.2} 
	\centering
	
	\caption{BadEncoder achieves high ASRs.}
	\begin{tabular}{|c|c|c|c|}
		\hline
	\makecell{Pre-training \\Dataset}& \makecell{ Target Downs-\\tream Dataset }& ASR-B (\%)	 & ASR (\%)  \\ \hline
	\multirow{3}{*}{CIFAR10}
	&	GTSRB  & 2.79 & 98.64  \\ \cline{2-4} 
	&	SVHN & 37.53 & 99.14  \\ \cline{2-4} 
	&	STL10 & 10.38 & 99.73  \\ \hline
	
	\multirow{3}{*}{STL10}
	&	GTSRB  & 1.67 & 95.04  \\ \cline{2-4} 
	&	SVHN  & 46.11 & 97.64  \\ \cline{2-4} 
	&	CIFAR10  & 12.30 & 98.51  \\ \hline
	\end{tabular}
	\label{highasr_table} 
	\vspace{-4mm}
\end{table}

\subsection{Experimental Results}
\vspace{-2mm}
\myparatight{BadEncoder achieves high ASRs} Table~\ref{highasr_table} shows the ASR-B without injecting backdoor to the pre-trained image encoder and ASR of BadEncoder. The experimental results indicate that BadEncoder can achieve high attack success rates. For instance, when the pre-training dataset is CIFAR10 and the target downstream dataset is STL10, BadEncoder can achieve 99.73\% attack success rate. In contrast, the attack success rate (i.e., ASR-B) is only  10.38\% when we do not inject backdoor to the pre-trained image encoder. We note that the ASR-B is relatively high when the target downstream dataset is SVHN because SVHN is unbalanced and the selected target class happens to be the most popular class. 

BadEncoder is successful because our backdoored image encoder outputs similar feature vectors for the reference input and the trigger-embedded inputs, which makes the backdoored downstream classifier predict the same class (i.e., target class) for them.  
Given a clean (or backdoored) image encoder, we use it to produce a feature vector for the reference input and each trigger-embedded testing input in the target downstream dataset, and we calculate the cosine similarity between the feature vector of the reference input and that of each trigger-embedded testing input. Figure~\ref{compare_similarity_of_clean_backdoor} in Appendix shows the cumulative distribution functions (CDFs) of such cosine similarity scores for the clean image encoder and backdoored image encoder. Our results show that our backdoored image encoder produces much more similar feature vectors for the reference input and the trigger-embedded inputs than the clean image encoder, which is the reason why our ASR is much higher than ASR-B.

\begin{table}[tp]\renewcommand{\arraystretch}{1.2} 
	\centering
	\caption{Our BadEncoder maintains the accuracy of the downstream classifiers. }
	\setlength{\tabcolsep}{1mm}
	{
	\begin{tabular}{|c|c|c|c|c|}
		\hline
	\makecell{Pre-training \\Dataset} & \makecell{Target  Downs-\\tream Dataset} & \makecell{Downstream \\Dataset} & CA (\%) & BA (\%)  \\ \hline
	\multirow{9}{*}{CIFAR10}
	&	\multirow{3}{*}{GTSRB} 
	    & GTSRB & 81.84 & 82.27 \\ \cline{3-5} 
	    && SVHN & 58.50 & 68.93  \\ \cline{3-5} 
	    && STL10 & 76.14 & 75.94  \\ \cline{2-5} 
	&	\multirow{3}{*}{SVHN} 
	    & GTSRB & 81.84 & 82.19  \\ \cline{3-5} 
	    &&	SVHN & 58.50 & 69.32 \\ \cline{3-5} 
	    && STL10 & 76.14 & 75.66  \\ \cline{2-5}   
	&	\multirow{3}{*}{STL10} 
	    & GTSRB & 81.84 & 82.55  \\ \cline{3-5}  
	    && SVHN & 58.50 & 68.68  \\ \cline{3-5} 
	    &&	STL10 & 76.14 & 76.18 \\ \hline
	    
	\multirow{9}{*}{STL10}
	&	\multirow{3}{*}{GTSRB} 
		&	GTSRB & 76.12 & 76.63  \\ \cline{3-5} 
	    && SVHN & 55.35 & 63.85  \\ \cline{3-5} 
	    && CIFAR10 & 86.77 & 86.63  \\ \cline{2-5} 
	&	\multirow{3}{*}{SVHN} 
	    & GTSRB & 76.12 & 75.45  \\ \cline{3-5} 
	    &&	SVHN & 55.35 & 65.59 \\ \cline{3-5} 
	    && CIFAR10 & 86.77 & 86.23 \\ \cline{2-5}   
	&	\multirow{3}{*}{CIFAR10} 
	    & GTSRB & 76.12 & 76.47  \\ \cline{3-5}  
	    && SVHN & 55.35 & 64.37 \\ \cline{3-5}  
	    &&	CIFAR10 & 86.77 & 86.55  \\ \hline
	\end{tabular}
	}
	\label{maintail_utility_table}
	\vspace{-5mm}
\end{table}

\myparatight{BadEncoder preserves accuracy of the downstream classifiers} Table~\ref{maintail_utility_table} compares the clean accuracy and the  backdoored accuracy in different scenarios. In particular, given a pre-training dataset and a target downstream dataset, we evaluate the clean accuracy of the downstream classifiers and  backdoored accuracy of the backdoored downstream classifiers for both the target downstream dataset and non-target downstream datasets. Our experimental results indicate that our BadEncoder preserves the accuracy of the downstream classifiers for the target/non-target downstream datasets. In particular,  
the differences between the backdoor accuracies and the clean accuracies are within 1\% in most cases. We note that the backdoor accuracies are higher than the clean accuracies when the downstream dataset is SVHN, which we consistently observe in all our experiments. We suspect the reason is that the SVHN dataset is noisy, i.e., many images in the SVHN dataset contain distracting digits,  
 and our backdoored image encoder produces better features for the SVHN dataset after fine-tuning the clean image encoder.

\myparatight{Impact of loss terms} Our BadEncoder leverages three loss terms, i.e., $L_0$, $L_1$, and $L_2$ in Equation~\ref{final_loss_term}. Moreover, we use $\lambda_1$ (or $\lambda_2$) to weight $L_1$ (or $L_2$). Therefore, we explore the impact of the three loss terms and the parameters $\lambda_1$ and $\lambda_2$ on our BadEncoder. 
Table~\ref{impact_of_loss_term} shows our experimental results when we exclude a loss term in our BadEncoder, where $\lambda_1=1$ (or $\lambda_2=1$) if $L_1$ (or $L_2$) is included. Our results show that all the three loss terms are necessary for our BadEncoder to achieve high attack success rates. Moreover,  $L_2$ is necessary to achieve a high  backdoored accuracy, i.e., to preserve the accuracy of the downstream classifier. This is because $L_2$ is designed to achieve the utility goal. Note that although $L_2$ is designed for the utility goal, excluding it also substantially reduces the attack success rate. This is because, without $L_2$, the backdoored downstream classifier becomes less accurate, in particular it misclassifies the reference input.

\begin{figure}[!t]
	 \centering
\vspace{-5mm}
\subfloat[$\lambda_1$]{\includegraphics[width=0.25\textwidth]{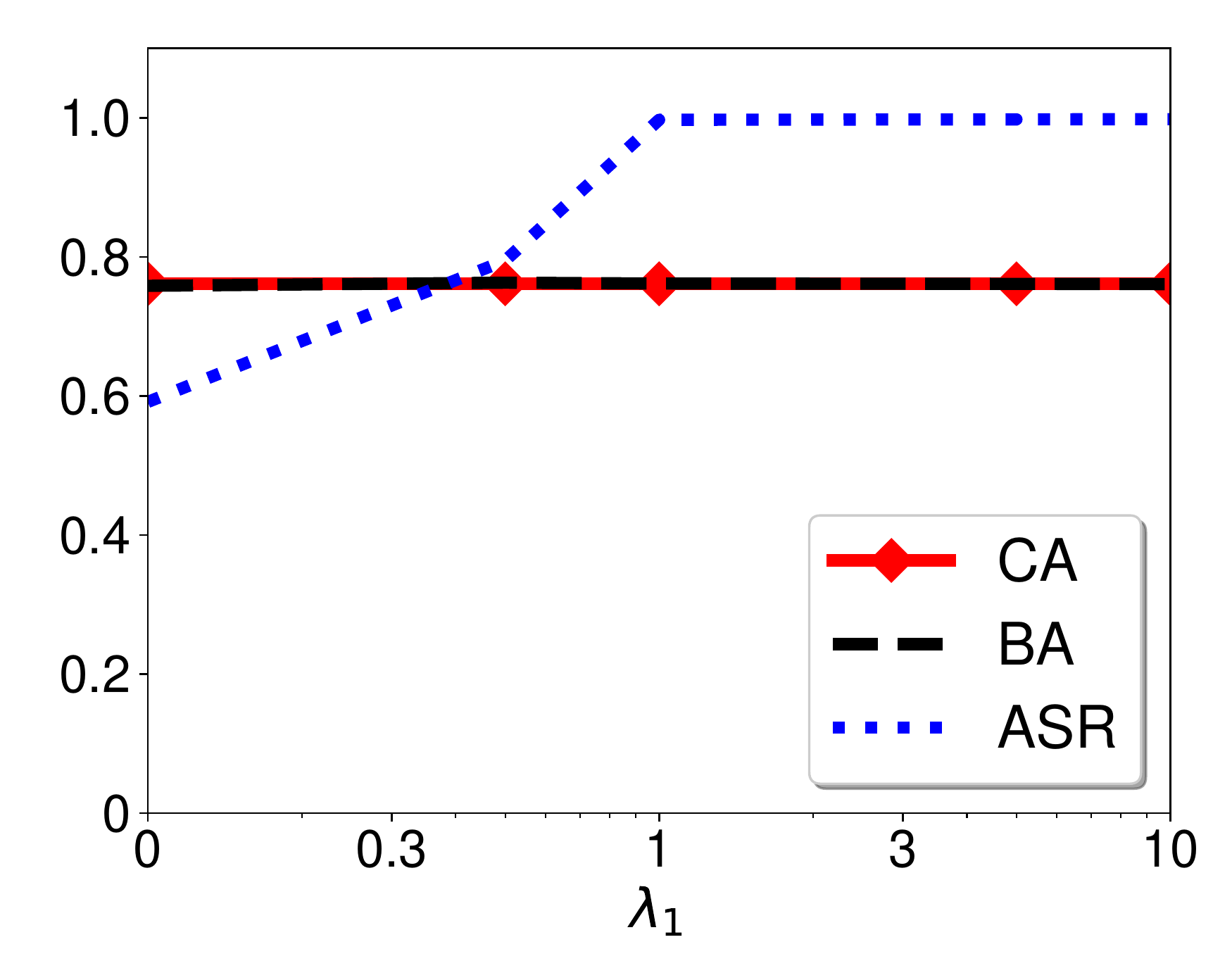}}
\subfloat[$\lambda_2$]{\includegraphics[width=0.25\textwidth]{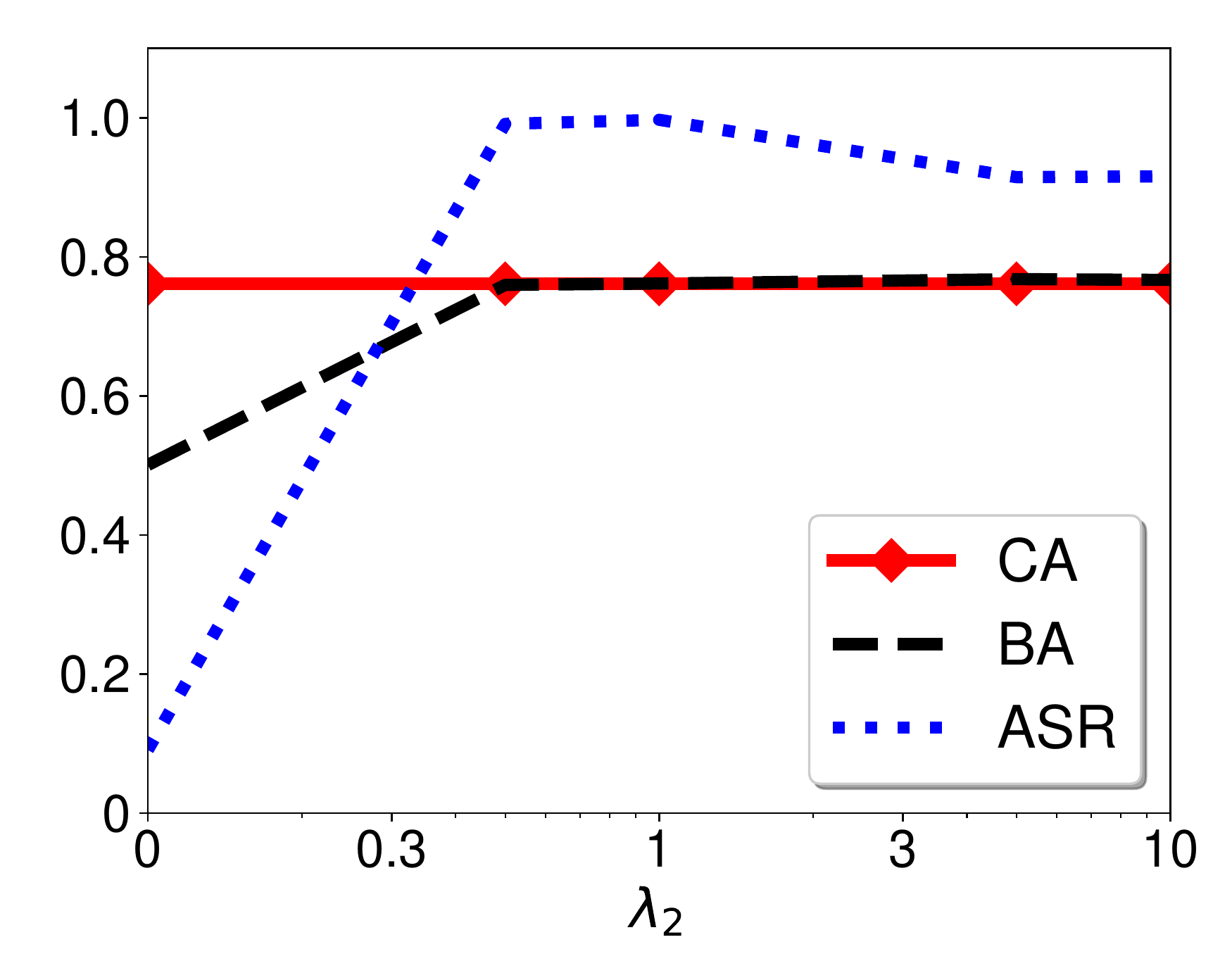}}
\caption{The impacts of $\lambda_1$ and $\lambda_2$.}
\label{impact_of_lambda1_lambda2}
\end{figure}

\begin{table}[tp]\renewcommand{\arraystretch}{1.2} 
	\centering
	\caption{The impact of the loss terms.} 
	\begin{tabular}{|c|c|c|c|}
		\hline
  \makecell{Removed \\Loss Terms} & CA(\%) & BA(\%) & ASR(\%)  \\ \hline
	
		$L_0$ & \multirow{4}{*}{76.14} & 76.48 & 9.48  \\ \cline{1-1} \cline{3-4} 
		$L_1$ &  & 75.85 &  59.15 \\ \cline{1-1} \cline{3-4} 
		$L_2$ &  & 50.08 &  9.09 \\ \cline{1-1} \cline{3-4} 
		None &  & 76.18 & 99.73   \\ \hline
	\end{tabular}
	\label{impact_of_loss_term}
	\vspace{-4mm}
\end{table}

We also study the impact of $\lambda_1$ and $\lambda_2$ on the backdoored accuracy and attack success rate of our BadEncoder. Figure~\ref{impact_of_lambda1_lambda2} shows the results. First, we observe that BadEncoder achieves high attack success rates and preserve accuracy after $\lambda_1$ and $\lambda_2$ are larger than some thresholds. Second, our BadEncoder is less sensitive to $\lambda_1$. In particular, the attack success rate starts to decrease after $\lambda_2$ is larger than around 1, while the attack success rate keeps stable even if  $\lambda_1$ is as large as 10. This is because our shadow dataset size (related to $L_2$) is much larger than the number of reference inputs (related to $L_1$).

\myparatight{Impact of shadow dataset} A shadow dataset can be characterized by its size and distribution. Therefore, we study the impact of both the size and distribution of the shadow dataset on our BadEncoder. 
Figure~\ref{impact_of_attack_dataset_size} and Figure~\ref{impact_of_attack_dataset_size_svhn_stl10} in Appendix show the impact of the shadow dataset size on BadEncoder. We find that BadEncoder achieves high attack success rates and preserves accuracy of the downstream classifiers once the shadow dataset size is larger than around 20\% of the pre-training dataset. We note that the  backdoored accuracy for the SVHN dataset increases as the size of the shadow dataset increases. We suspect the reason is that SVHN is noisy and our backdoored image encoder produces more distinguishable features for it after fine-tuning the clean image encoder.

We also study the impact of the shadow dataset distribution on our BadEncoder. In particular, we consider three cases. In the first case, the shadow dataset is a subset of the pre-training dataset. In particular,  we randomly sample $10,000$ images from the pre-training dataset CIFAR10 as the shadow dataset. In the second case,  the shadow dataset has the same distribution as the pre-training dataset but does not  overlap with it. In particular, we use the testing $10,000$ images of CIFAR10, which were not used to pre-train the image encoder, as the shadow dataset. In the third case,  the shadow dataset has a different distribution with the pre-training dataset. Specifically, we randomly sample 10,000 images from the Food101 dataset~\cite{bossard14} which contains images of food and  has a different distribution with CIFAR10.
Table~\ref{impact_of_attack_dist} shows our experimental results. Our results show that our BadEncoder achieves high attack success rates and preserves accuracy of the downstream classifiers in all the three cases, though the attack success rates are lower in the third case when the shadow dataset has a different distribution with the pre-training dataset. 
Our results indicate that our shadow dataset does not need to be from the pre-training dataset nor follow its  distribution.

\begin{figure}[!t]
	 \centering
	 \vspace{-5mm}
\subfloat[]{\includegraphics[width=0.25\textwidth]{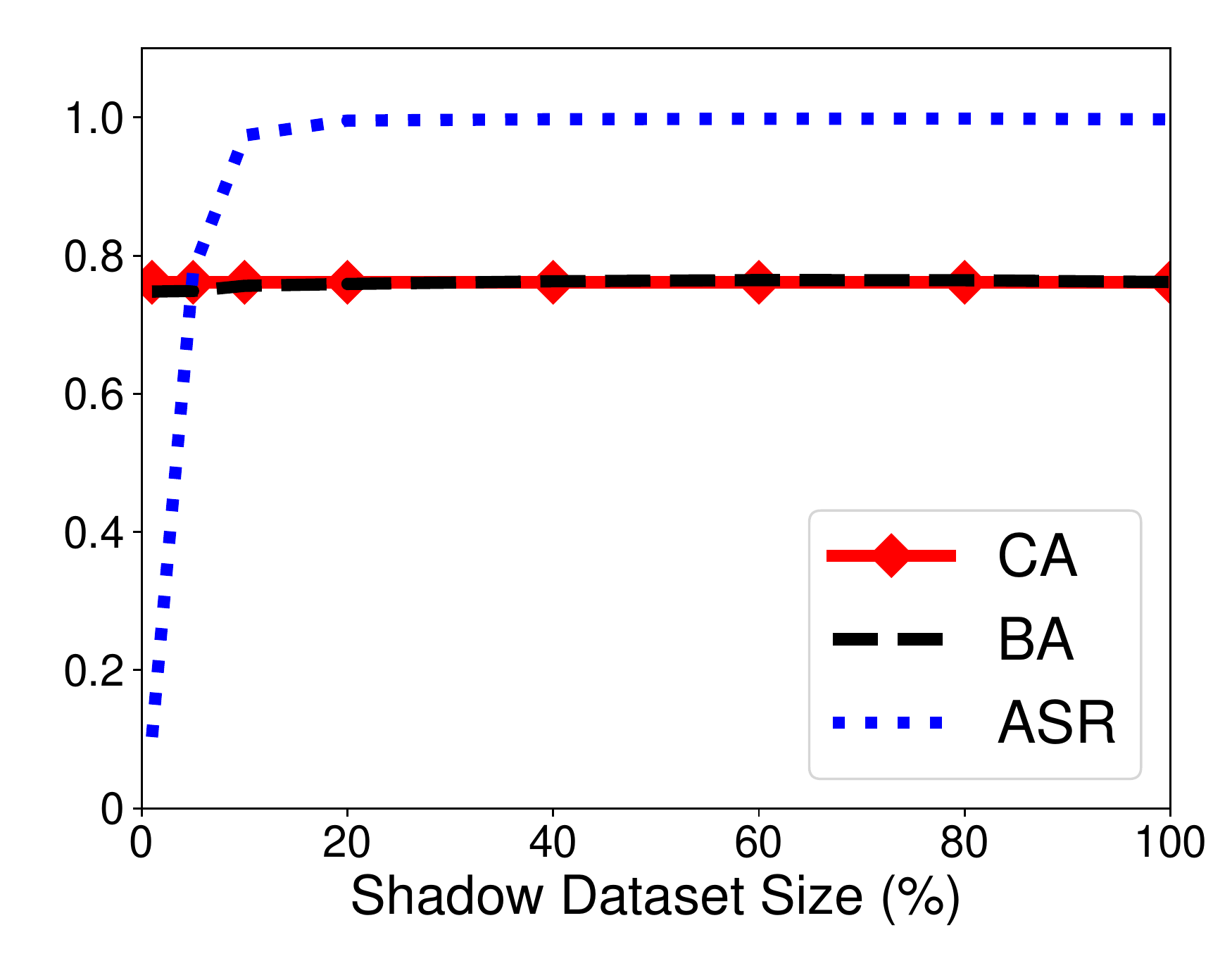}\label{impact_of_attack_dataset_size}}
\subfloat[]{\includegraphics[width=0.25\textwidth]{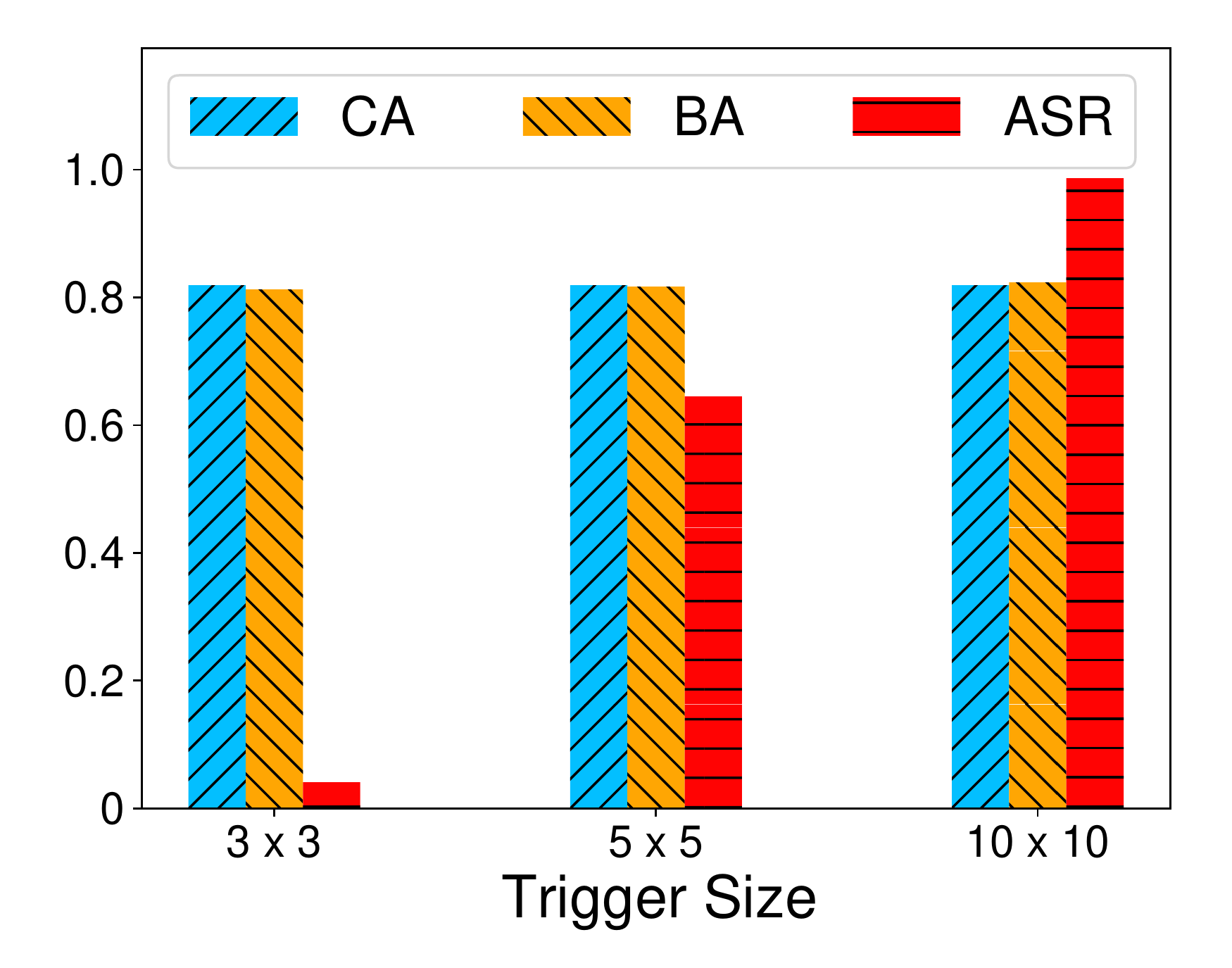}\label{impact_of_trigger_size}
\vspace{-3mm}
}
\caption{(a) The impact of the shadow dataset size on  BadEncoder when  the target downstream dataset is STL10 and the shadow dataset is a subset of the pre-training dataset. (b) The impact of the trigger size on BadEncoder when the target downstream dataset is GTSRB. }
\vspace{-2mm}
\end{figure}

\begin{table}[tp]\renewcommand{\arraystretch}{1.4} 
\setlength{\tabcolsep}{3pt}
	\centering
	\fontsize{7.5}{8}\selectfont
	\caption{The impact of the shadow dataset's distribution on  BadEncoder.} 
	\begin{tabular}{|c|c|c|c|c|}
		\hline
	 \makecell{Target Downs-\\tream Dataset}& \makecell{Shadow Dataset} & CA (\%) & BA (\%) & ASR (\%)   \\ \hline
    
	\multirow{3}{*}{GTSRB}
	&	A subset of pre-training dataset & \multirow{3}{*}{81.84} & 81.21 & 98.19  \\ \cline{2-2}\cline{4-5} 
	&	Same distribution & & 81.12 & 97.52  \\ \cline{2-2}\cline{4-5} 
	&	Different distributions & & 82.21 & 93.27  \\ \cline{1-3}\cline{4-5}

	\multirow{3}{*}{SVHN}
	&		A subset of pre-training dataset  & \multirow{3}{*}{ 58.50} & 62.32 & 98.30  \\ \cline{2-2}\cline{4-5} 
	&	Same distribution &  & 62.07 & 98.06  \\ \cline{2-2}\cline{4-5} 
	& Different distributions &  & 60.40 & 84.80  \\ \cline{1-5} 

    \multirow{3}{*}{STL10}
	&		A subset of pre-training dataset  & \multirow{3}{*}{76.14} & 75.90 & 99.55  \\ \cline{2-2}\cline{4-5} 
	&	Same distribution &  & 75.70 & 99.43  \\ \cline{2-2}\cline{4-5} 
	& Different distributions & & 75.99 & 98.15  \\ \hline

	\end{tabular}
	\label{impact_of_attack_dist}
	\vspace{-4mm}
\end{table}

\myparatight{Impact of trigger size} Figure~\ref{impact_of_trigger_size},  Figure~\ref{impact_of_trigger_size_cifar10} (in Appendix), and Figure~\ref{impact_of_trigger_size_stl10} (in Appendix)  show the impact of the trigger size on BadEncoder. Note that our trigger is a white square located at the bottom right corner of an input and trigger size refers to the height/width of a trigger. We have the following observations from our experimental results. First, our BadEncoder achieves high attack success rates when the trigger size is no smaller than some threshold, e.g., $10 \times 10$, $3 \times 3$, and $5 \times 5$ respectively for GTSRB,  SVHN, and STL10 when the pre-training dataset is CIFAR10. We also observe that such threshold depends on both the pre-training dataset and the target downstream dataset. For instance, the GTSRB downstream dataset requires $10 \times 10$ and $3 \times 3$ trigger sizes to achieve high attack success rates when the pre-training dataset is CIFAR10 and STL10, respectively. 
Second, the backdoor accuracies are comparable to (or higher than) the clean accuracies when the trigger has different sizes. In other words, our BadEncoder with different trigger sizes do not sacrifice the utility of the pre-trained image encoder.

\myparatight{Multiple reference inputs} Our BadEncoder relies on that  the backdoored downstream classifier correctly predicts the   reference input as the target class. However, it is possible that the backdoored downstream classifier does not correctly predict the reference input into the target class. In response, the attacker can use multiple reference inputs for a target class. Table~\ref{multiple_attack_inputs} shows our  results when our BadEncoder uses each of three reference inputs and all the three reference inputs, where the three reference inputs were collected from the Internet and are shown in Figure~\ref{multiple_classes_attack_inputs} in Appendix. As the results show, one reference input (Truck 0) has high attack success rate because it is correctly classified by the backdoored downstream classifier, but  two reference inputs (Truck 1 and Truck 2) lead to low attack success rates because they are misclassified by the backdoored downstream classifier. 
However,  our BadEncoder  can still achieve high attack success rate when using all the three reference inputs. In other words, our BadEncoder is effective once at least one reference input is correctly classified by the backdoored downstream classifier. 

\myparatight{Multiple target classes} Our BadEncoder can  attack multiple target classes in a target downstream task simultaneously. In particular, we use a different trigger for each target class. Specifically, we select ``airplane, ``truck'', and ``horse'' as the three target classes, and we use $10 \times 10$ white square (each pixel has value (255, 255, 255)), green square (each pixel has value (0, 255, 0)), and purple square (each pixel has value (255, 0, 255)) located at the bottom right corner, upper left corner, and central of an image as the corresponding triggers. We collected an reference input for each target class from the Internet, and the reference inputs are shown in Figure~\ref{multiple_downstream_inputs} in Appendix.   Table~\ref{multiple_target_classes} shows our  results. We find that our BadEncoder can still achieve high attack success rates while maintaining the accuracy of the downstream classifier when attacking multiple target classes simultaneously.

\begin{table}[tp]\renewcommand{\arraystretch}{1.2} 
	\centering
	\caption{Results of using multiple reference inputs. } 
	\begin{tabular}{|c|c|c|c|c|c|}
		\hline
	 Reference Input & CA (\%) & BA (\%) & ASR (\%)   \\ \hline
		Truck 0 & \multirow{4}{*}{76.14} & 76.18 & 99.73 \\ \cline{1-1}  \cline{3-4} 
		Truck 1 & & 76.10 & 11.64  \\ \cline{1-1}  \cline{3-4} 
		Truck 2 &  & 75.63 & 0.23 \\ \cline{1-1}  \cline{3-4} 
		All &  & 75.33 & 99.24  \\ \cline{1-4} 
	\end{tabular}
	\label{multiple_attack_inputs}
\end{table}

\begin{table}[!t]\renewcommand{\arraystretch}{1.2} 
	\centering
	\caption{Results of attacking three target classes simultaneously. } 
	\begin{tabular}{|c|c|c|c|}
		\hline
	 Target Class & CA (\%) & BA (\%) & ASR (\%)   \\ \hline

		Airplane & \multirow{3}{*}{76.14} & \multirow{3}{*}{75.61} & 99.73  \\ \cline{1-1}\cline{4-4} 
		Horse &  &  & 99.86  \\ \cline{1-1} \cline{4-4} 
		Truck & & & 100.00  \\ \hline
	\end{tabular}
	\label{multiple_target_classes}
	\vspace{-1mm}
\end{table}

\myparatight{Multiple target downstream tasks} 
An attacker may be interested in attacking multiple target downstream tasks simultaneously. To evaluate our attacks in such scenario, we use CIFAR10 as the pre-training dataset and simultaneously attack the other three datasets as the target downstream datasets. We adopt the default target class and reference input for each target downstream dataset. 
Moreover, we adopt a $10 \times 10$ white square, green square, and purple square located at the bottom right corner, upper left corner, and central of an image as the corresponding triggers for the three target downstream datasets, respectively. Table~\ref{multiple_downstream_tasks} shows our results. Our results show that our attacks can achieve high attack success rates at attacking multiple target downstream tasks simultaneously while maintaining accuracy of the downstream classifiers.

\begin{table}[tp]\renewcommand{\arraystretch}{1.2} 
	\centering
	\caption{Results of attacking three target downstream datasets simultaneously. } 
	\begin{tabular}{|c|c|c|c|c|}
		\hline
	\makecell{Target Downstream Dataset} & CA (\%) & BA (\%) & ASR (\%)   \\ \hline

		GTSRB & 81.84 & 82.98 & 93.35  \\ \cline{1-4} 
		SVHN & 58.50 & 69.74 & 100.00  \\ \cline{1-4} 
		STL10 & 76.14 & 75.75 & 100.00  \\ \hline
	\end{tabular}
	\label{multiple_downstream_tasks}
	\vspace{-5mm}
\end{table}

 \myparatight{Impact of other parameters} We also studied the impact of other parameters (e.g., learning rate, number of epochs, trigger value) on BadEncoder. We found that BadEncoder achieves high ASRs while maintaining accuracy for the downstream classifier across different parameter settings. Due to the limited space, the results are shown in Table~\ref{result_of_different_parameters} in Appendix.

\myparatight{Comparing BadEncoder with Latent Backdoor Attack (LBA)~\cite{yao2019latent}} We compare BadEncoder with LBA, which was designed for transfer learning. When extended to self-supervised learning, LBA tries to inject a backdoor into a teacher classifier trained based on an image encoder. When the backdoored teacher classifier is fine-tuned to train a student classifier (i.e., downstream classifier in our terminology) for a target downstream task, the student classifier inherits the backdoor. To train a backdoored teacher classifier, LBA requires a large \emph{labeled} dataset consisting of {labeled} examples in both the target class and  non-target classes from the target downstream task. 
Specifically, we use the same reference input in the target class in BadEncoder for LBA. Moreover, we further randomly sample a certain fraction of the testing images in the non-target classes of the target downstream dataset for LBA. We adopt a public implementation from the authors of LBA~\cite{latent_url}.

Table~\ref{compare_with_LBA} in Appendix shows our comparison results when different fractions of testing images in the non-target classes of the target downstream dataset are used by LBA. Our experimental results indicate that BadEncoder achieves higher attack success rates than LBA, where the attack success rates are evaluated using the rest of testing images of the target downstream dataset that were not used by LBA. The reason is that BadEncoder injects the backdoor via directly optimizing the pre-trained image encoder while LBA injects a backdoor into a teacher classifier built based on the image encoder.

\section{Two Real-world Case Studies}
We show two real-world case studies for our BadEncoder. In particular, we apply our BadEncoder to an image encoder pre-trained on ImageNet and released by Google~\cite{chen2020simple}. Moreover, we apply our BadEncoder to CLIP~\cite{radford2021learning}, which includes an image encoder and a text encoder pre-trained on 400 million (image, text) pairs collected from the Internet. CLIP was released by OpenAI~\cite{clip_url}. Since CLIP includes a text encoder, it can be used for zero-shot classifier.

\subsection{Attacking Image Encoder Pre-trained on ImageNet}
\subsubsection{Experimental Setup} 
\label{experimental_setup_imagenet}
We consider the attacker selects a single target downstream task/dataset, a single target class, and a single reference input. In particular, we select ``truck'', ``priority sign'', and ``digit one'' as the target classes for the datasets STL10, GTSRB, and SVHN, respectively. Similarly, we collected the reference input for each target class from the Internet, which were shown in Figure~\ref{resnet50_google_attack_input} in Appendix. We set $\lambda_1 = 1$ and $\lambda_2 = 1$; we randomly sample $1\%$ of the training images of ImageNet as the shadow dataset; we use a $50 \times 50$ white square located at the bottom right corner of an image as the trigger. Moreover, we adopt the same neural network architecture in Section~\ref{sec:exp} as the downstream classifiers. We note that each image in ImageNet  was resized to $224 \times 224 \times 3$ when Google used them to pre-train the image encoder. Therefore, we also resize each image in the shadow dataset and downstream datasets to be $224 \times 224 \times 3$ in our experiments.  We fine-tune the pre-trained image encoder for 200 epochs with learning rate $10^{-4}$ and batch size $16$ to inject the backdoor. Note that we use a small batch size due to the large resolution of images in ImageNet.

\subsubsection{Experimental Results} Table~\ref{case_study_resnet50_google} shows the experimental results. We find that our BadEncoder can achieve high attack success rates while maintaining the accuracy of the downstream classifiers. Our experimental results demonstrate that our BadEncoder is effective when applied to an image encoder that is pre-trained on a large amount of unlabeled images. 

\begin{table}[tp]\renewcommand{\arraystretch}{1.2} 
	\centering
	\caption{BadEncoder achieves high attack success rates and maintains the accuracy of the downstream classifiers when attacking the image encoder pre-trained on ImageNet by Google~\cite{chen2020simple}. }
	\begin{tabular}{|c|c|c|c|c|}
		\hline
 \makecell{Target Downs-\\tream Dataset} & CA (\%) & BA (\%) & ASR-B (\%)	 & ASR (\%)  \\ \hline
		GTSRB & 76.53 & 78.42 & 5.47 & 98.93  \\ \cline{1-5} 
		STL10 & 95.66 & 95.68 & 10.24 & 99.99  \\ \cline{1-5} 
		SVHN & 72.55 & 73.77 & 32.28 & 99.93 \\  \hline
	\end{tabular}
	\label{case_study_resnet50_google}
	\vspace{-5mm}
\end{table}

\begin{table}[tp]\renewcommand{\arraystretch}{1.2} 
	\centering
	\caption{ BadEncoder achieves high attack success rates and maintains the  accuracy of the downstream classifiers when attacking CLIP~\cite{radford2021learning}.}
	\subfloat[Multi-shot classifiers]{
	\begin{tabular}{|c|c|c|c|c|}
		\hline
 \makecell{Target Downs-\\tream Dataset} & CA (\%) & BA (\%) & ASR-B (\%)	 & ASR (\%)   \\ \hline
		GTSRB & 82.36 & 82.14 & 5.37 & 99.33  \\ \cline{1-5} 
		STL10 & 97.09 & 96.69 & 10.00 & 99.81  \\ \cline{1-5} 
		SVHN & 70.60 & 70.27 & 20.79 & 99.99  \\ \hline
	\end{tabular}\label{case_study_clip_st_1}}
	
		\subfloat[Zero-shot classifiers]{
		\begin{tabular}{|c|c|c|c|c|}
		\hline
	 \makecell{Target Downs-\\tream Dataset} & CA (\%) & BA (\%) & ASR-B (\%)	 & ASR (\%)   \\ \hline

		GTSRB & 29.83 & 29.84 & 1.96 & 99.82  \\ \cline{1-5} 
		STL10 & 94.60 & 92.80 & 10.08 & 99.96  \\ \cline{1-5} 
		SVHN & 11.73 & 11.16 & 53.55 & 100.00  \\ \hline
	\end{tabular}
	\label{case_study_clip_zs}}

	\label{case_study_clip_st}
	\vspace{-6mm}
\end{table}

\subsection{Attacking CLIP}

\subsubsection{Experimental Setup} CLIP consists of both an image encoder and a text encoder. We apply BadEncoder to inject a backdoor to the image encoder. When building a downstream classifier, CLIP supports both multi-shot classifier and zero-shot classifier, as we discussed in Section~\ref{sec:background}. Therefore, we evaluate BadEncoder for both scenarios. Since we don't have access to CLIP's  pre-training dataset, we  adopt the training images of CIFAR10 as the shadow dataset. In both scenarios, we fine-tune the CLIP's image encoder for 200 epochs using our Algorithm~\ref{alg:example} with learning rate $10^{-6}$ and batch size $16$.  

In the multi-shot classifier scenario, we consider  the same experimental settings as  those when attacking the image encoder pre-trained on ImageNet (please refer to Section~\ref{experimental_setup_imagenet} for details). 
Moreover, we collected the reference inputs  from the Internet and they can be found in Figure~\ref{resnet50_clip_attack_input} in Appendix. 
In the zero-shot classifier scenario, we also consider a single target downstream dataset and a target class. We select ``truck'', ``stop sign'', and ``digit zero'' as target classes for the target downstream datasets STL10, GTSRB, and SVHN, respectively. We collected a reference input for each target class from the Internet and they are shown in Figure~\ref{clip_zero_shot_attack_input} in Appendix. Recall that a zero-shot classifier requires a context sentence for each class. We adopt the context sentences ``A photo of a \{class name\}'' for STL10 and SVHN. However, for GTSRB, we adopt the context sentences ``A traffic sign photo of a \{class name\}'' because we found they achieve better accuracy than ``A photo of a \{class name\}'' for GTSRB.

\subsubsection{Experimental Results} Table~\ref{case_study_clip_st} shows the experimental results. We find that our BadEncoder achieves high attack success rates and maintains the  accuracy of the downstream classifiers (both multi-shot classifiers and zero-shot classifiers). Our experimental results indicate that our BadEncoder is effective when applied to an image encoder pre-trained on a large amount of (image, text) pairs.

\section{Defenses}

\begin{table}[tp]\renewcommand{\arraystretch}{1.2} 
	\centering
	
	\caption{Anomaly Indices for the backdoored downstream classifiers produced by Neural Cleanse~\cite{wang2019neural}. A classifier is predicted to be backdoored if the Anomaly Index is larger than 2.  The pre-training dataset is CIFAR10. }
	\begin{tabular}{|c|c|c|}
		\hline
	\makecell{ Target Downstream Dataset }&  \makecell{Anomaly Index}	  \\ \hline

		GTSRB  & 1.940   \\ \cline{1-2} 
		SVHN & 1.340  \\ \cline{1-2} 
		STL10 & 1.251   \\ \hline
	
	\end{tabular}
	\label{neural_cleanse_ai} 
	\vspace{-5mm}
\end{table}

Many defenses~\cite{tran2018spectral,liu2018fine,gao2019strip,xu2019detecting,liu2019abs,wang2019neural,doan2020februus,chiang2019certified,levine2020randomized,xiang2020patchguard,metzen2021efficient} were proposed to defend classifiers against backdoor attacks. In particular, we can categorize these defenses into two types: \emph{empirical defenses}~\cite{liu2018fine,gao2019strip,xu2019detecting,liu2019abs,wang2019neural,doan2020februus} and \emph{provable defenses}~\cite{chiang2019certified,levine2020randomized,xiang2020patchguard,wang2020certifying,zhang2020backdoor,metzen2021efficient}. We evaluate our attack against two state-of-the-art empirical defenses (i.e., Neural Cleanse~\cite{wang2019neural} and MNTD~\cite{xu2019detecting}) and a state-of-the-art provable defense (i.e., PatchGuard~\cite{xiang2020patchguard}). Moreover, we generalize MNTD to detect backdoored encoders. Our experimental results indicate that  Neural Cleanse, MNTD, and generalized MNTD cannot detect our  attack, while PatchGuard provides insufficient certified robustness guarantees under our attack. 

\subsection{Neural Cleanse} 

Neural Cleanse~\cite{wang2019neural} aims to detect whether a classifier (i.e., a downstream classifier in our context) is backdoored or not. In particular, they first try to reverse engineer a trigger for each possible class  and then use anomaly detection to predict whether the classifier is backdoored or not. 
Specifically,  Neural Cleanse produces an \emph{Anomaly Index} for  a given classifier. The classifier is predicted to be backdoored if the Anomaly Index is larger than 2. We consider our attacks with the default parameter settings in Section~\ref{sec:exp}, e.g., the pre-training dataset is CIFAR10, a single target downstream dataset, a single target class, and a single reference input. 
We use Neural Cleanse to detect backdoor in a backdoored downstream classifier. Note that we cannot directly apply Neural Cleanse to the  image encoder because  it is designed for classifiers. We use a public implementation~\cite{neural_cleanse} released by the authors, and adopt their suggested parameter settings. We note that  Neural Cleanse requires a clean dataset. We use the testing dataset of a target downstream dataset as a clean dataset in our evaluation. Table~\ref{neural_cleanse_ai} shows the Anomaly Index produced by Neural Cleanse for each backdoored downstream classifier trained based on our backdoored image encoder. We find that the Anomaly Indices are consistently smaller than 2. In other words, Neural Cleanse cannot detect our backdoor attacks in the backdoored downstream classifiers.

\subsection{MNTD}
MNTD~\cite{xu2019detecting} aims to detect whether a classifier is backdoored or not using a binary meta-classifier. Roughly speaking, MNTD first trains a set of clean and backdoored shadow classifiers, then extracts a feature representation for each shadow classifier, and finally trains a binary meta-classifier based on the feature representations. We use \CR{jumbo MNTD with query-tuning} to detect backdoored downstream classifiers and also extend it to detect backdoored encoders. We perform our experiments in our default parameter setting, e.g., we adopt CIFAR10 as the pre-training dataset and STL10 as the target downstream dataset.

\noindent
{\bf Detecting backdoored downstream classifiers:} We first use CIFAR10 to pre-train a clean image encoder and then use BadEncoder to craft a backdoored image encoder. We respectively use the clean image encoder and backdoored image encoder to train 10 clean downstream classifiers and 10 backdoored downstream classifiers on STL10. We aim to use MNTD to classify these 20 downstream classifiers to be backdoored or not. 

We adopt the publicly available code of MNTD in our implementation~\cite{mntd_url}.  We use the training dataset of STL10 to train each shadow classifier and we train \CR{200} clean shadow classifiers in total, where the architecture of a shadow classifier is the composed architecture of the image encoder and the downstream classifier. \CR{Following~\cite{xu2019detecting}, we adopt jumbo learning to train \CR{200} backdoored shadow classifiers on the training dataset of STL10.}  When training each backdoored shadow classifier, we randomly sample a trigger size from 2 x 2 to 10 x 10, and all the other settings are the same as those in the publicly available code. Based on the \CR{200} clean shadow classifiers and \CR{200} backdoored shadow classifiers, we train \CR{50 sequential meta-classifiers and report their average detection accuracy.} \CR{Note that we adopt the query-tuning technique to find the best set of inputs (following~\cite{xu2019detecting})}, i.e., the meta-classifiers and the inputs to construct feature representations of shadow classifiers are jointly optimized. 

Then, we apply the meta-classifiers to classify the 20 downstream classifiers. In particular, we use the clean (or backdoored) encoder to compute the feature vectors of the inputs; then we calculate the outputs of a clean (or backdoored) downstream classifier for the inputs’ feature vectors; and finally a meta-classifier predicts the downstream classifier to be backdoored or not based on the concatenated outputs. The
average detection accuracy of the \CR{50 meta-classifiers is 0.5 (random guessing) and the standard deviation is 0.00}, indicating the ineffectiveness of MNTD against BadEncoder. We suspect the reason is that BadEncoder does not compromise the training of downstream classifiers. 

\noindent
{\bf Detecting backdoored encoders:}
We generalize \CR{jumbo MNTD with query-tuning}~\cite{xu2019detecting} to detect backdoored encoders. Our idea is to treat an encoder as if it was a classifier. Following~\cite{xu2019detecting}, we assume the defender has access to 2\% of the pre-training dataset. The defender (e.g., a downstream customer) pre-trains \CR{200} clean shadow encoders using the same setting as described in Section~\ref{pre_training_image_encoder}. For each clean shadow encoder, the defender crafts a backdoored shadow encoder using our BadEncoder algorithm. Specifically, the defender samples an input from the training dataset of its downstream dataset as the reference input, generates a random trigger whose size is randomly sampled from 2 x 2 to 10 x 10, and treats its accessible pre-training dataset as the shadow dataset. The defender concatenates the feature vectors produced by a shadow encoder for a set of inputs as a feature representation of the shadow encoder. Given the feature representations of the \CR{200} clean and \CR{200} backdoored shadow encoders, the defender trains a binary meta-classifier. Following~\cite{xu2019detecting}, we jointly optimize the meta-classifier and the set of inputs. Moreover, we also train \CR{50 sequential meta-classifiers and report their average detection accuracy.}

Given an encoder, we first extract its feature representation using the set of optimized inputs and then use a meta-classifier to classify it to be backdoored or clean. We train three clean encoders on CIFAR10 using different initializations and craft three backdoored encoders using our BadEncoder algorithm accordingly. The average detection accuracy of the \CR{50} meta-classifiers for the six encoders is \CR{0.52 and the standard deviation is 0.17}.\footnote{\CR{One possible reason for the large standard deviation is that we only have six encoders in detection and thus the detection accuracy of each meta-classifier can only be 0, 0.17, 0.33, 0.5, 0.67, 0.83, and 1.}} Our results show that MNTD is \CR{slightly} more accurate at detecting backdoored encoders than backdoored downstream classifiers under our attacks, but the average detection accuracy is still low.

We note that Xu et al.~\cite{xu2019detecting} showed 64 clean and backdoored shadow classifiers are sufficient to train an accurate meta-classifier. We trained 200 clean shadow classifiers (or encoders) and 200 backdoored shadow classifiers (or encoders). However, we acknowledge that the defender may achieve higher detection accuracy by training more clean and backdoored shadow classifiers (or encoders), e.g., Xu et al. explored up to 2,048 clean shadow classifiers and 2,048 backdoored shadow classifiers.

\subsection{PatchGuard}
BadEncoder adds a trigger (i.e., a patch) to an input image such that a backdoored downstream classifier classifies the trigger-embedded input into the target class. Therefore, we can leverage provable defenses~\cite{chiang2019certified,levine2020randomized,xiang2020patchguard,metzen2021efficient} against adversarial patches to mitigate BadEncoder. Given an input, such a defense provably guarantees that the predicted label is unaffected by the trigger/patch once its size is no larger than a threshold. Moreover, given a testing dataset and a trigger size, such a defense can produce a lower bound of accuracy no matter what trigger/patch is embedded to a testing input once its size is no larger than the given trigger size. The lower bound of accuracy is known as \emph{certified accuracy}.  
Among such defenses, PatchGuard~\cite{xiang2020patchguard} achieves the state-of-the-art certified accuracy. Specifically, PatchGuard is based on two insights. First, an adversarial patch/trigger can only corrupt a small number of extracted features of a convolutional neural network (CNN) with a small receptive field for an input. Second, the robust aggregation of extracted features can limit the impact of the small number of corrupted features. PatchGuard designs a robust feature aggregation algorithm, namely \emph{robust masking}, which is compatible with any CNN with small receptive fields.

We evaluate BadEncoder against PatchGuard. Specifically, BadEncoder uses the default parameter settings in Section~\ref{sec:exp}, and we use PatchGuard (i.e., the variant with Mask-DS) to defend the backdoored downstream classifiers. 
We adopt a public implementation~\cite{patchguard_url} of PatchGuard.
Table~\ref{patchguard_evaluation} shows the certified accuracy and attack success rates for the backdoored downstream classifiers. We find that PatchGuard is insufficient for defending against BadEncoder. Specifically, although our attack success rates decrease, the certified accuracy is all 0. 
We suspect the reason is that PatchGuard’s certified accuracy is a loose lower bound of accuracy. In particular, PatchGuard considers embedding a (different) trigger to each testing input independently, while the same trigger is embedded to testing inputs in backdoor attacks.

\begin{table}[tp]\renewcommand{\arraystretch}{1.2} 
	\centering
	
	\caption{The  certified accuracy and attack success rates for the backdoored downstream classifiers defended by PatchGuard~\cite{xiang2020patchguard}. The pre-training dataset is CIFAR10. }
	\begin{tabular}{|c|c|c|c|}
		\hline
 \makecell{ Target Downstream Dataset }&  \makecell{Certified Accuracy (\%)} & ASR (\%)	  \\ \hline
	
		GTSRB  & 0 & 56.34 \\ \cline{1-3} 
		SVHN & 0  & 59.89\\ \cline{1-3} 
		STL10 & 0  & 46.46 \\ \hline
	
	\end{tabular}
	\label{patchguard_evaluation} 
	\vspace{-5mm}
\end{table}

\section{Related Work}

\subsection{Self-supervised Learning}
Self-supervised learning is a new AI paradigm that aims to pre-train encoders that can be used for many downstream tasks using a large amount of unlabeled data. It has been applied to a variety of domains such as  natural language processing (NLP), graph, and computer vision, and has achieved state-of-the-art performance in many downstream tasks in these domains. 
Specifically, in the   NLP domain, many pre-trained language models~\cite{devlin2018bert,radford2018improving,radford2019language,brown2020language,yang2019xlnet} were proposed. Specifically, the idea is to pre-train a language model on a large amount of unlabeled text. The pre-trained language model can be further used for many downstream NLP tasks such as text classification and question answering. In the graph domain, self-supervised learning has been used to pre-train Graph Neural Networks (GNNs)~\cite{hu2020strategies,qiu2020gcc} to learn  transferable structural graph representations. The pre-trained GNN can be used for many downstream tasks, e.g., graph classification. 
In the computer vision domain, an image encoder can be pre-trained using unlabeled images~\cite{hadsell2006dimensionality,he2020momentum,chen2020simple,hjelm2018learning,grill2020bootstrap} or  (image, text) pairs~\cite{srivastava2012multimodal,joulin2016learning,thomee2016yfcc100m,li2020unicoder,radford2021learning}. When   (image, text) pairs are used, a text encoder is also pre-trained and can be used for zero-shot classification.

\subsection{Backdoor Attacks}
Deep neural networks are vulnerable to backdoor attacks in various domains~\cite{gu2017badnets,chen2017targeted,liutrojaning2018,dai2019backdoor,bagdasaryan2020backdoor,zhang2020backdoor,xi2020graph}, e.g., image, text, and graph. In this work, we focus on the image domain. Next, we review backdoor attacks in these domains.

\noindent
{\bf Image:} In backdoor attacks to image classification~\cite{gu2017badnets,chen2017targeted,liao2018backdoor,liutrojaning2018,yao2019latent,saha2020hidden,turner2019label,li2019invisible,tan2019bypassing,liu2020reflection,salem2020dynamic}, an attacker aims to inject a hidden behavior 
into a classifier. In particular, the backdoored classifier predicts any image embedded with a trigger into a target class. Existing backdoor attacks~\cite{gu2017badnets,chen2017targeted,liutrojaning2018,yao2019latent} directly inject a backdoor into a classifier. For instance,  BadNets~\cite{gu2017badnets} injects a backdoor into a classifier via poisoning its training images, i.e., adding a backdoor trigger to the training inputs and changing their labels to the target class. Liu et al.~\cite{liutrojaning2018} proposed to first invert a classifier to generate a backdoor trigger, and then inject the backdoor into the classifier via retraining it.

Yao et al.~\cite{yao2019latent} proposed latent backdoor attack (LBA) to transfer learning. When extended to self-supervised learning,  LBA injects a backdoor into a teacher classifier built based on the image encoder and a labeled dataset similar to the target downstream dataset. When the backdoored teacher classifier is used to fine-tune a student classifier for the target downstream task, the student classifier inherits the backdoor behavior. The key differences between LBA and our BadEncoder are as follows. First, LBA requires a large labeled dataset for the target downstream task. 
In contrast, our BadEncoder only requires a few reference inputs from the target class and an arbitrary unlabeled shadow dataset. Second, as demonstrated by our experiments, even if such labeled data is available, LBA achieves suboptimal attack effectiveness when extended to self-supervised learning.  
Third, our BadEncoder can attack multiple target downstream tasks simultaneously while LBA only attacks a single target downstream task by design.

We note that Carlini et al.~\cite{carlini2021poisoning} recently proposed data poisoning and backdoor attacks to self-supervised learning, which is concurrent to ours. Their attacks poison the pre-training dataset, but they were designed for (image, text) pairs based self-supervised learning.

\noindent
{\bf Text:} Some studies~\cite{dai2019backdoor,chen2020badnl,zhang2020trojaning} showed that  natural language classifiers are also vulnerable to backdoor attacks.  
For instance, Zhang et al.~\cite{zhang2020trojaning} proposed backdoor attacks to pre-trained language models. These studies are different from ours as we focus on image encoders. Moreover, Zhang et al. require an attacker to have substantial knowledge about the downstream tasks. For instance, an attacker requires access to a subset of the training dataset of the target downstream task. In contrast, our attack does not require such information.

\noindent
{\bf Graph:} Backdoor attacks have also been studied for graph classification~\cite{zhang2020backdoor,xi2020graph}. For instance, Zhang et al.~\cite{zhang2020backdoor} developed a subgraph based backdoor attack to  GNNs. 
Xi et al.~\cite{xi2020graph} also proposed a subgraph based backdoor attack, which considers both topological structures and descriptive features when designing backdoor triggers.

We note that some studies~\cite{ji2017backdoor,ji2018model,shafahi2018poison}
proposed targeted data poisoning attacks that involve a feature extractor (i.e., pre-trained image encoder in our context). For instance, Ji et al.~\cite{ji2018model} modified a feature extractor such that a downstream classifier predicts a particular attacker-chosen clean target input as attacker-chosen target class. Shafahi et al.~\cite{shafahi2018poison} adds perturbation to some training inputs in the target class in the target downstream dataset such that the clean feature extractor produces similar feature vectors for the perturbed training inputs and the attacker-chosen clean target inputs; and then the downstream classifier trained using the perturbed training inputs will predict the clean target inputs as the target class. The key difference with our work is that our attack injects backdoors into the image encoder such that a backdoored downstream classifier predicts the target class for \emph{any} input embedded with a pre-defined trigger.

\subsection{Defenses against Backdoor Attacks}
 Defenses against backdoor attacks can be categorized into \emph{empirical defenses}~\cite{tran2018spectral,wang2019neural,guo2019tabor,chen2018detecting,xu2019detecting,gao2019strip,liu2019abs,chen2019deepinspect,tang2019demon,chou2020sentinet,doan2020februus} and \emph{provable defenses}~\cite{chiang2019certified,levine2020randomized,xiang2020patchguard,wang2020certifying,weber2020rab,zhang2020backdoor,metzen2021efficient}. 
 
\noindent
{\bf Empirical defenses:} A family of empirical defenses~\cite{tran2018spectral,wang2019neural,xu2019detecting,gao2019strip,liu2019abs,doan2020februus} 
aim to detect whether there is a backdoor (or a trigger) in a classifier (or an input). For instance, Wang et al.~\cite{wang2019neural} proposed Neural Cleanse~\cite{wang2019neural} which tries to detect whether a classifier (i.e., a downstream classifier in our context) is backdoored or not. In particular, they first try to reverse engineer a trigger for each possible class  and then use anomaly detection to predict whether the classifier is backdoored or not. Liu et al.~\cite{liu2019abs} proposed to detect  backdoor via analyzing the behaviors of a neuron under different levels of stimulation. Gao et al.~\cite{gao2019strip} proposed STRIP, which predicts an  input has a  trigger embedded if the predicted labels for randomly perturbed versions of the input have small entropy. 
Another family of empirical defenses~\cite{liu2018fine,wang2019neural} try to remove the backdoor in a classifier. For instance, Liu et al.~\cite{liu2018fine} proposed to first prune the neurons that are less informative and then fine-tune the pruned classifier to remove the backdoor. Our results show that \CRR{Neural Cleanse and MNTD, two state-of-the-art empirical defenses, cannot detect our backdoor attacks.}

\noindent
{\bf Provable defenses:} Provable defenses~\cite{wang2020certifying,weber2020rab,zhang2020backdoor,metzen2021efficient,chiang2019certified,levine2020randomized,xiang2020patchguard,jia2020certified,jia2021intrinsic} can provide provable robustness guarantees against backdoor attacks. In particular, given an input, these defenses can provably guarantee that the predicted label remains unchanged when the trigger size is smaller than a threshold. For instance, Wang et al.~\cite{wang2020certifying} leveraged randomized smoothing to mitigate backdoor attacks and found that existing randomized smoothing techniques provide limited provable robustness guarantees.
Since backdoor attacks embed a trigger/patch to a testing input, provable defenses against adversarial patches can be used to mitigate them.
Chiang et al.~\cite{chiang2019certified} proposed the first provable defense against adversarial patches, which leverages interval bound propagation. 
Recently, Xiang et al.~\cite{xiang2020patchguard} proposed PatchGuard which achieves the state-of-the-art certified accuracy against adversarial patches. However, our experimental results indicate that PatchGuard provides insufficient  robustness guarantees under our attacks. 

\section{Conclusion and Future Work}
In this work, we show that pre-trained image encoders in 
self-supervised learning are vulnerable to backdoor attacks. 
Injecting  backdoors to image encoders can be formulated as an optimization problem, which can be solved by a gradient descent based method. 
We also find that existing defenses against backdoor attacks are insufficient to defend against our attack. 
Interesting future work includes: 1) generalizing our attack to self-supervised learning in other domains, e.g., natural language processing and graph, 2) developing new defenses to defend against our attack, and 3) studying how to pre-train an encoder (assume the encoder is clean) such that the downstream classifiers built based on the encoder are more robust against conventional backdoor attacks that compromise the training of downstream classifiers~\cite{gu2017badnets,chen2017targeted,liutrojaning2018}.

\section*{Acknowledgements}
We thank the anonymous reviewers and our shepherd Fabio Pierazzi for constructive comments. This work was supported by  NSF under Grant No. 1937786 and
the Army Research Office under Grant No. W911NF2110182. 

\bibliographystyle{IEEEtran}
\bibliography{refs}

\begin{thebibliography}{10}
\providecommand{\url}[1]{#1}
\csname url@samestyle\endcsname
\providecommand{\newblock}{\relax}
\providecommand{\bibinfo}[2]{#2}
\providecommand{\BIBentrySTDinterwordspacing}{\spaceskip=0pt\relax}
\providecommand{\BIBentryALTinterwordstretchfactor}{4}
\providecommand{\BIBentryALTinterwordspacing}{\spaceskip=\fontdimen2\font plus
\BIBentryALTinterwordstretchfactor\fontdimen3\font minus
  \fontdimen4\font\relax}
\providecommand{\BIBforeignlanguage}[2]{{%
\expandafter\ifx\csname l@#1\endcsname\relax
\typeout{** WARNING: IEEEtran.bst: No hyphenation pattern has been}%
\typeout{** loaded for the language `#1'. Using the pattern for}%
\typeout{** the default language instead.}%
\else
\language=\csname l@#1\endcsname
\fi
#2}}
\providecommand{\BIBdecl}{\relax}
\BIBdecl

\bibitem{devlin2018bert}
J.~Devlin, M.-W. Chang, K.~Lee, and K.~Toutanova, ``Bert: Pre-training of deep
  bidirectional transformers for language understanding,'' in \emph{NAACL},
  2019.

\bibitem{hadsell2006dimensionality}
R.~Hadsell, S.~Chopra, and Y.~LeCun, ``Dimensionality reduction by learning an
  invariant mapping,'' in \emph{CVPR}, 2006.

\bibitem{he2020momentum}
K.~He, H.~Fan, Y.~Wu, S.~Xie, and R.~Girshick, ``Momentum contrast for
  unsupervised visual representation learning,'' in \emph{CVPR}, 2020.

\bibitem{chen2020simple}
T.~Chen, S.~Kornblith, M.~Norouzi, and G.~Hinton, ``A simple framework for
  contrastive learning of visual representations,'' in \emph{ICML}, 2020.

\bibitem{hjelm2018learning}
R.~D. Hjelm, A.~Fedorov, S.~Lavoie-Marchildon, K.~Grewal, P.~Bachman,
  A.~Trischler, and Y.~Bengio, ``Learning deep representations by mutual
  information estimation and maximization,'' in \emph{ICLR}, 2019.

\bibitem{grill2020bootstrap}
J.-B. Grill, F.~Strub, F.~Altch{\'e}, C.~Tallec, P.~H. Richemond,
  E.~Buchatskaya, C.~Doersch, B.~A. Pires, Z.~D. Guo, M.~G. Azar \emph{et~al.},
  ``Bootstrap your own latent: A new approach to self-supervised learning,'' in
  \emph{NeurIPS}, 2020.

\bibitem{radford2021learning}
A.~Radford, J.~W. Kim, C.~Hallacy, A.~Ramesh, G.~Goh, S.~Agarwal, G.~Sastry,
  A.~Askell, P.~Mishkin, J.~Clark \emph{et~al.}, ``Learning transferable visual
  models from natural language supervision,'' \emph{arXiv}, 2021.

\bibitem{gu2017badnets}
T.~Gu, B.~Dolan-Gavitt, and S.~Garg, ``Badnets: Identifying vulnerabilities in
  the machine learning model supply chain,'' \emph{IEEE Access}, 2017.

\bibitem{chen2017targeted}
X.~Chen, C.~Liu, B.~Li, K.~Lu, and D.~Song, ``Targeted backdoor attacks on deep
  learning systems using data poisoning,'' \emph{arXiv preprint
  arXiv:1712.05526}, 2017.

\bibitem{liutrojaning2018}
Y.~Liu, S.~Ma, Y.~Aafer, W.-C. Lee, J.~Zhai, W.~Wang, and X.~Zhang, ``Trojaning
  attack on neural networks,'' in \emph{NDSS}, 2018.

\bibitem{bagdasaryan2020blind}
E.~Bagdasaryan and V.~Shmatikov, ``Blind backdoors in deep learning models,''
  in \emph{Usenix Security}, 2021.

\bibitem{yao2019latent}
Y.~Yao, H.~Li, H.~Zheng, and B.~Y. Zhao, ``Latent backdoor attacks on deep
  neural networks,'' in \emph{CCS}, 2019.

\bibitem{wang2019neural}
B.~Wang, Y.~Yao, S.~Shan, H.~Li, B.~Viswanath, H.~Zheng, and B.~Y. Zhao,
  ``Neural cleanse: Identifying and mitigating backdoor attacks in neural
  networks,'' in \emph{IEEE S\& P}, 2019.

\bibitem{xu2019detecting}
X.~Xu, Q.~Wang, H.~Li, N.~Borisov, C.~A. Gunter, and B.~Li, ``Detecting ai
  trojans using meta neural analysis,'' in \emph{IEEE S \& P}, 2021.

\bibitem{xiang2020patchguard}
C.~Xiang, A.~N. Bhagoji, V.~Sehwag, and P.~Mittal, ``Patchguard: Provable
  defense against adversarial patches using masks on small receptive fields,''
  in \emph{Usenix Security}, 2021.

\bibitem{pathak2016context}
D.~Pathak, P.~Krahenbuhl, J.~Donahue, T.~Darrell, and A.~A. Efros, ``Context
  encoders: Feature learning by inpainting,'' in \emph{CVPR}, 2016.

\bibitem{noroozi2016unsupervised}
M.~Noroozi and P.~Favaro, ``Unsupervised learning of visual representations by
  solving jigsaw puzzles,'' in \emph{ECCV}, 2016.

\bibitem{srivastava2012multimodal}
N.~Srivastava, R.~Salakhutdinov \emph{et~al.}, ``Multimodal learning with deep
  boltzmann machines.'' in \emph{NeurIPS}, 2012.

\bibitem{joulin2016learning}
A.~Joulin, L.~Van Der~Maaten, A.~Jabri, and N.~Vasilache, ``Learning visual
  features from large weakly supervised data,'' in \emph{ECCV}, 2016.

\bibitem{thomee2016yfcc100m}
B.~Thomee, D.~A. Shamma, G.~Friedland, B.~Elizalde, K.~Ni, D.~Poland, D.~Borth,
  and L.-J. Li, ``Yfcc100m: The new data in multimedia research,''
  \emph{Communications of the ACM}, vol.~59, no.~2, pp. 64--73, 2016.

\bibitem{li2020unicoder}
G.~Li, N.~Duan, Y.~Fang, M.~Gong, and D.~Jiang, ``Unicoder-vl: A universal
  encoder for vision and language by cross-modal pre-training,'' in
  \emph{AAAI}, 2020.

\bibitem{krizhevsky2009learning}
A.~Krizhevsky, ``Learning multiple layers of features from tiny images,''
  \emph{Tech Report}, 2009.

\bibitem{coates2011analysis}
A.~Coates, A.~Ng, and H.~Lee, ``An analysis of single-layer networks in
  unsupervised feature learning,'' in \emph{AISTATS}, 2011.

\bibitem{stallkamp2012man}
J.~Stallkamp, M.~Schlipsing, J.~Salmen, and C.~Igel, ``Man vs. computer:
  Benchmarking machine learning algorithms for traffic sign recognition,''
  \emph{Neural networks}, vol.~32, pp. 323--332, 2012.

\bibitem{netzer2011reading}
Y.~Netzer, T.~Wang, A.~Coates, A.~Bissacco, B.~Wu, and A.~Y. Ng, ``Reading
  digits in natural images with unsupervised feature learning,'' in \emph{NIPS
  Workshop on Deep Learning and Unsupervised Feature Learning}, 2011.

\bibitem{bossard14}
L.~Bossard, M.~Guillaumin, and L.~Van~Gool, ``Food-101 -- mining discriminative
  components with random forests,'' in \emph{ECCV}, 2014.

\bibitem{he2016deep}
K.~He, X.~Zhang, S.~Ren, and J.~Sun, ``Deep residual learning for image
  recognition,'' in \emph{CVPR}, 2016.

\bibitem{simclr_url}
``{SimCLR},'' \url{https://github.com/google-research/simclr}.

\bibitem{simclr_url_pytorch}
``{SimCLR PyTorch},'' \url{https://github.com/leftthomas/SimCLR}.

\bibitem{latent_url}
``{LBA},'' \url{http://sandlab.cs.uchicago.edu/latent/}.

\bibitem{clip_url}
``{CLIP},'' \url{https://github.com/openai/CLIP}.

\bibitem{tran2018spectral}
B.~Tran, J.~Li, and A.~Madry, ``Spectral signatures in backdoor attacks,'' in
  \emph{NeurIPS}, 2018.

\bibitem{liu2018fine}
K.~Liu, B.~Dolan-Gavitt, and S.~Garg, ``Fine-pruning: Defending against
  backdooring attacks on deep neural networks,'' in \emph{RAID}, 2018.

\bibitem{gao2019strip}
Y.~Gao, C.~Xu, D.~Wang, S.~Chen, D.~C. Ranasinghe, and S.~Nepal, ``Strip: A
  defence against trojan attacks on deep neural networks,'' in \emph{ACSAC},
  2019.

\bibitem{liu2019abs}
Y.~Liu, W.-C. Lee, G.~Tao, S.~Ma, Y.~Aafer, and X.~Zhang, ``Abs: Scanning
  neural networks for back-doors by artificial brain stimulation,'' in
  \emph{CCS}, 2019.

\bibitem{doan2020februus}
B.~G. Doan, E.~Abbasnejad, and D.~C. Ranasinghe, ``Februus: Input purification
  defense against trojan attacks on deep neural network systems,'' in
  \emph{ACSAC}, 2020.

\bibitem{chiang2019certified}
P.-y. Chiang, R.~Ni, A.~Abdelkader, C.~Zhu, C.~Studor, and T.~Goldstein,
  ``Certified defenses for adversarial patches,'' in \emph{ICLR}, 2019.

\bibitem{levine2020randomized}
A.~Levine and S.~Feizi, ``(de) randomized smoothing for certifiable defense
  against patch attacks,'' in \emph{NeurIPS}, 2020.

\bibitem{metzen2021efficient}
J.~H. Metzen and M.~Yatsura, ``Efficient certified defenses against patch
  attacks on image classifiers,'' in \emph{ICLR}, 2021.

\bibitem{wang2020certifying}
B.~Wang, X.~Cao, J.~Jia, N.~Z. Gong \emph{et~al.}, ``On certifying robustness
  against backdoor attacks via randomized smoothing,'' in \emph{CVPR 2020
  Workshop on Adversarial Machine Learning in Computer Vision}, 2020.

\bibitem{zhang2020backdoor}
Z.~Zhang, J.~Jia, B.~Wang, and N.~Z. Gong, ``Backdoor attacks to graph neural
  networks,'' in \emph{SACMAT}, 2021.

\bibitem{neural_cleanse}
``{Neural Cleanse},'' \url{https://github.com/bolunwang/backdoor}.

\bibitem{mntd_url}
``{MNTD},'' \url{https://github.com/AI-secure/Meta-Nerual-Trojan-Detection}.

\bibitem{patchguard_url}
``{PatchGuard},'' \url{https://github.com/inspire-group/PatchGuard}.

\bibitem{radford2018improving}
\BIBentryALTinterwordspacing
A.~Radford, K.~Narasimhan, T.~Salimans, and I.~Sutskever, ``Improving language
  understanding by generative pre-training,'' 2018. [Online]. Available:
  \url{https://s3-us-west-2.amazonaws.com/openai-assets/research-covers/language-unsupervised/language_understanding_paper.pdf}
\BIBentrySTDinterwordspacing

\bibitem{radford2019language}
A.~Radford, J.~Wu, R.~Child, D.~Luan, D.~Amodei, and I.~Sutskever, ``Language
  models are unsupervised multitask learners,'' \emph{OpenAI blog}, vol.~1,
  no.~8, p.~9, 2019.

\bibitem{brown2020language}
T.~B. Brown, B.~Mann, N.~Ryder, M.~Subbiah, J.~Kaplan, P.~Dhariwal,
  A.~Neelakantan, P.~Shyam, G.~Sastry, A.~Askell \emph{et~al.}, ``Language
  models are few-shot learners,'' \emph{Arxiv:2005.14165}, 2020.

\bibitem{yang2019xlnet}
Z.~Yang, Z.~Dai, Y.~Yang, J.~Carbonell, R.~R. Salakhutdinov, and Q.~V. Le,
  ``Xlnet: Generalized autoregressive pretraining for language understanding,''
  \emph{NeurIPS}, 2019.

\bibitem{hu2020strategies}
W.~Hu, B.~Liu, J.~Gomes, M.~Zitnik, P.~Liang, V.~Pande, and J.~Leskovec,
  ``Strategies for pre-training graph neural networks,'' in \emph{ICLR}, 2020.

\bibitem{qiu2020gcc}
J.~Qiu, Q.~Chen, Y.~Dong, J.~Zhang, H.~Yang, M.~Ding, K.~Wang, and J.~Tang,
  ``Gcc: Graph contrastive coding for graph neural network pre-training,'' in
  \emph{KDD}, 2020.

\bibitem{dai2019backdoor}
J.~Dai, C.~Chen, and Y.~Li, ``A backdoor attack against lstm-based text
  classification systems,'' \emph{IEEE Access}, vol.~7, pp. 138\,872--138\,878,
  2019.

\bibitem{bagdasaryan2020backdoor}
E.~Bagdasaryan, A.~Veit, Y.~Hua, D.~Estrin, and V.~Shmatikov, ``How to backdoor
  federated learning,'' in \emph{AISTATS}, 2020.

\bibitem{xi2020graph}
Z.~Xi, R.~Pang, S.~Ji, and T.~Wang, ``Graph backdoor,'' in \emph{Usenix
  Security}, 2021.

\bibitem{liao2018backdoor}
C.~Liao, H.~Zhong, A.~Squicciarini, S.~Zhu, and D.~Miller, ``Backdoor embedding
  in convolutional neural network models via invisible perturbation,'' in
  \emph{CODASPY}, 2020.

\bibitem{saha2020hidden}
A.~Saha, A.~Subramanya, and H.~Pirsiavash, ``Hidden trigger backdoor attacks,''
  in \emph{AAAI}, 2020.

\bibitem{turner2019label}
A.~Turner, D.~Tsipras, and A.~Madry, ``Label-consistent backdoor attacks,''
  \emph{arXiv preprint arXiv:1912.02771}, 2019.

\bibitem{li2019invisible}
S.~Li, B.~Z.~H. Zhao, J.~Yu, M.~Xue, D.~Kaafar, and H.~Zhu, ``Invisible
  backdoor attacks against deep neural networks,'' \emph{arXiv preprint
  arXiv:1909.02742}, 2019.

\bibitem{tan2019bypassing}
T.~J.~L. Tan and R.~Shokri, ``Bypassing backdoor detection algorithms in deep
  learning,'' in \emph{EuroS\&P}, 2020.

\bibitem{liu2020reflection}
Y.~Liu, X.~Ma, J.~Bailey, and F.~Lu, ``Reflection backdoor: A natural backdoor
  attack on deep neural networks,'' in \emph{ECCV}, 2020.

\bibitem{salem2020dynamic}
A.~Salem, R.~Wen, M.~Backes, S.~Ma, and Y.~Zhang, ``Dynamic backdoor attacks
  against machine learning models,'' \emph{arXiv preprint arXiv:2003.03675},
  2020.

\bibitem{carlini2021poisoning}
N.~Carlini and A.~Terzis, ``Poisoning and backdooring contrastive learning,''
  \emph{arXiv preprint arXiv:2106.09667}, 2021.

\bibitem{chen2020badnl}
X.~Chen, A.~Salem, M.~Backes, S.~Ma, and Y.~Zhang, ``Badnl: Backdoor attacks
  against nlp models,'' \emph{arXiv preprint arXiv:2006.01043}, 2020.

\bibitem{zhang2020trojaning}
X.~Zhang, Z.~Zhang, S.~Ji, and T.~Wang, ``Trojaning language models for fun and
  profit,'' in \emph{EuroS\&P}, 2021.

\bibitem{ji2017backdoor}
Y.~Ji, X.~Zhang, and T.~Wang, ``Backdoor attacks against learning systems,'' in
  \emph{CNS}, 2017.

\bibitem{ji2018model}
Y.~Ji, X.~Zhang, S.~Ji, X.~Luo, and T.~Wang, ``Model-reuse attacks on deep
  learning systems,'' in \emph{CCS}, 2018.

\bibitem{shafahi2018poison}
A.~Shafahi, W.~R. Huang, M.~Najibi, O.~Suciu, C.~Studer, T.~Dumitras, and
  T.~Goldstein, ``Poison frogs! targeted clean-label poisoning attacks on
  neural networks,'' in \emph{NeurIPS}, 2018.

\bibitem{guo2019tabor}
W.~Guo, L.~Wang, X.~Xing, M.~Du, and D.~Song, ``Tabor: A highly accurate
  approach to inspecting and restoring trojan backdoors in ai systems,''
  \emph{arXiv preprint arXiv:1908.01763}, 2019.

\bibitem{chen2018detecting}
B.~Chen, W.~Carvalho, N.~Baracaldo, H.~Ludwig, B.~Edwards, T.~Lee, I.~Molloy,
  and B.~Srivastava, ``Detecting backdoor attacks on deep neural networks by
  activation clustering,'' in \emph{AAAI}, 2019.

\bibitem{chen2019deepinspect}
H.~Chen, C.~Fu, J.~Zhao, and F.~Koushanfar, ``Deepinspect: A black-box trojan
  detection and mitigation framework for deep neural networks.'' in
  \emph{IJCAI}, 2019.

\bibitem{tang2019demon}
D.~Tang, X.~Wang, H.~Tang, and K.~Zhang, ``Demon in the variant: Statistical
  analysis of dnns for robust backdoor contamination detection,'' in
  \emph{Usenix Security}, 2021.

\bibitem{chou2020sentinet}
E.~Chou, F.~Tram{\`e}r, and G.~Pellegrino, ``Sentinet: Detecting localized
  universal attacks against deep learning systems,'' in \emph{2020 IEEE
  Security and Privacy Workshops (SPW)}.\hskip 1em plus 0.5em minus 0.4em\relax
  IEEE, 2020, pp. 48--54.

\bibitem{weber2020rab}
M.~Weber, X.~Xu, B.~Karlas, C.~Zhang, and B.~Li, ``Rab: Provable robustness
  against backdoor attacks,'' \emph{arXiv preprint arXiv:2003.08904}, 2020.

\bibitem{jia2020certified}
J.~Jia, X.~Cao, and N.~Z. Gong, ``Certified robustness of nearest neighbors
  against data poisoning attacks,'' \emph{arXiv preprint arXiv:2012.03765},
  2020.

\bibitem{jia2021intrinsic}
------, ``Intrinsic certified robustness of bagging against data poisoning
  attacks,'' in \emph{AAAI}, 2021.

\end{thebibliography}


\appendices

\begin{algorithm}[tb]
   \caption{BadEncoder}
   \label{alg:example}
\begin{algorithmic}
   \STATE {\bfseries Input:} $\theta$ (model parameters of $f$), $max\_epoch$, $lr$ (learning rate), $bs$ (batch size), $\mathcal{D}_{s}$, $\{\mathcal{R}_i\}_{i=1}^t$, and $\{\mathbf{e}_i\}_{i=1}^t$.
   \STATE {\bfseries Output:}  $\theta'$ (model parameters of $f'$). \\
   $\theta' \gets \theta$ \\
   $epoch \gets 1$ \\
   \WHILE  {$epoch \leq max\_epoch$}
   \STATE $iter \gets 1$ \\
   \WHILE{$iter \leq \lfloor |\mathcal{D}_s|/bs \rfloor$}
   \STATE  $\text{batch} \gets \textsc{MiniBatch}(\mathcal{D}_s, bs, step)$ \\
   \STATE $\theta' = \theta' - lr \cdot \nabla_{\theta'} L(\theta',\text{batch},\{\mathcal{R}_i\}_{i=1}^t, \{\mathbf{e}_i\}_{i=1}^t)$\\
   \STATE $iter \gets iter +1$ \\
   \ENDWHILE
   \STATE $epoch \gets epoch+1$ \\
   \ENDWHILE 
   \STATE \textbf{return} $\theta'$

\end{algorithmic}
\end{algorithm}

\begin{figure}[!t]
	 \centering
	\vspace{-2mm}
{\includegraphics[width=0.25\textwidth]{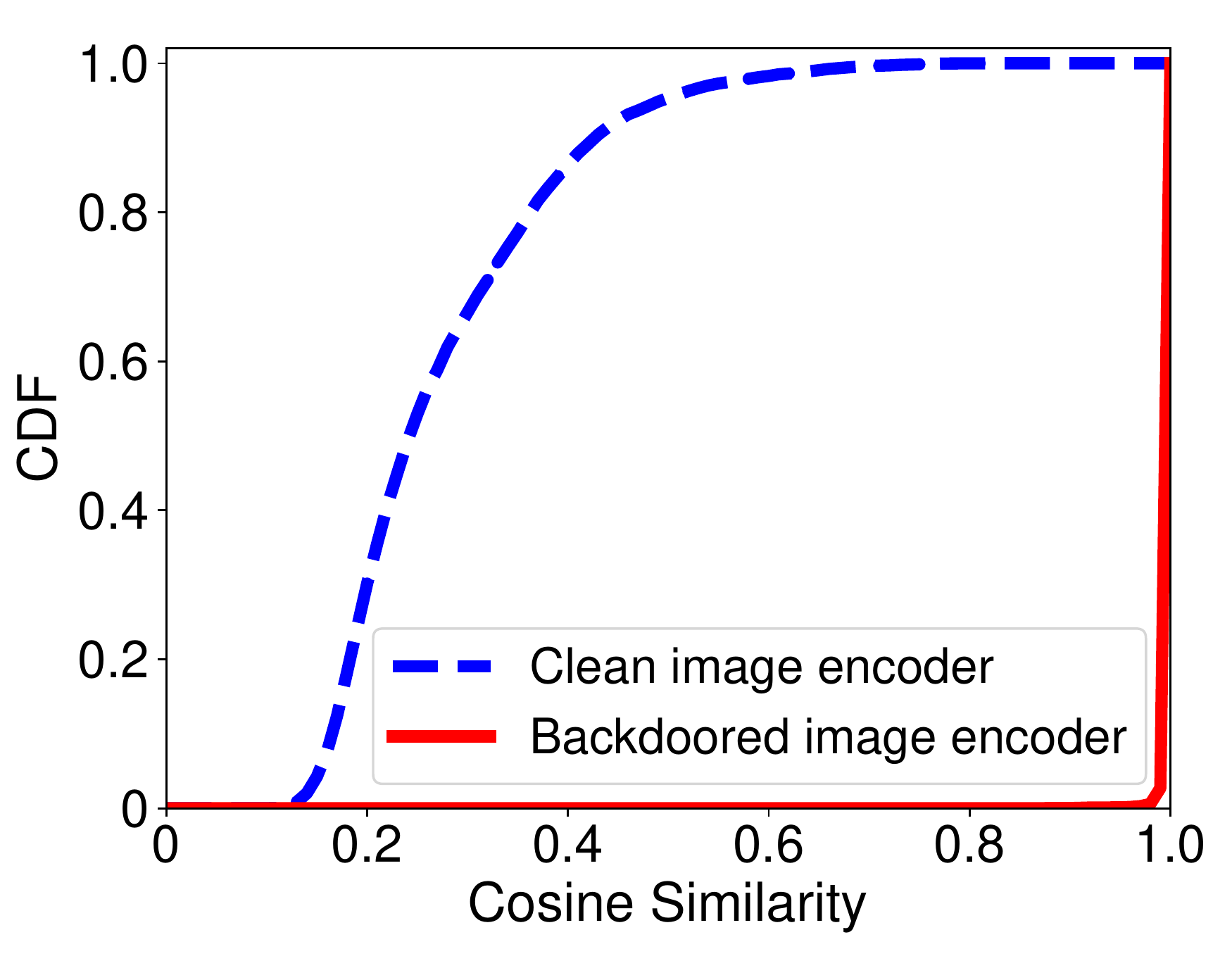}}
\vspace{-2mm}
\caption{The cumulative distribution functions (CDFs) of the cosine similarity scores between the feature vector of the reference input and those of the trigger-embedded inputs produced  by  the  clean image encoder and backdoored image encoder.} 
\label{compare_similarity_of_clean_backdoor}
\end{figure}

\begin{table}[!t]\renewcommand{\arraystretch}{1.2}
  \centering
  \caption{Comparing BadEncoder with LBA~\cite{yao2019latent}.} 
  \begin{tabular}{|c|c|c|c|}
    \hline
    \multirow{3}{*}{\makecell{Target Downs-\\tream Dataset}} &
      \multicolumn{2}{c|}{LBA} & 
      BadEncoder \cr\cline{2-4}
    & \makecell{ Fraction of Tes-\\ting Images (\%) } & ASR (\%) & ASR (\%) \\
    \hline
    \multirow{2}{*}{GTSRB} 
    & 10 & 7.14 & 98.73 \\ \cline{2-4}
    & 50 & 12.63 & 98.81 \\ \hline
    \multirow{2}{*}{SVHN}
    & 10 & 0.19 & 99.15 \\ \cline{2-4}
    & 50 & 24.16 & 99.09 \\ \hline
    \multirow{2}{*}{STL10}
    & 10 & 9.63 & 99.39 \\ \cline{2-4}
    & 50 & 12.77 & 99.59 \\ \hline
  \end{tabular}
  \label{compare_with_LBA}
\end{table}

\begin{table}[tp]\renewcommand{\arraystretch}{1.2} 
	\centering
	\caption{Results of using different parameter settings to pre-train an image encoder, craft a backdoored image encoder, and train a downstream classifier.}
	\setlength{\tabcolsep}{0.6mm}
	{
	\begin{tabular}{|c|c|c|c|c|c|}
		\hline
	\makecell{Stage} & \makecell{Parameter} & \makecell{Value} & CA (\%) & BA (\%) & ASR (\%)  \\ \hline
	\multirow{6}{*}{\makecell{Pre-training \\ an image \\encoder}}
	&	\multirow{3}{*}{\#epochs} 
	    & 500 & 74.61 & 75.11 & 99.66 \\ \cline{3-6} 
	    && 1,000 & 76.14 & 76.18 & 99.73  \\ \cline{3-6} 
	    && 1,500 & 75.95 & 76.65 & 99.86  \\ \cline{2-6} 
	&	\multirow{3}{*}{Learning rate} 
	    & $1\times 10^{-3}$ & 76.14 & 76.18 & 99.73  \\ \cline{3-6}
	    && $5\times 10^{-4}$ & 73.41 & 75.90 & 99.93  \\ \cline{3-6}  
	    &&	$1 \times 10^{-4}$ & 74.58 & 75.74 & 99.89 \\ \hline

	\multirow{8}{*}{\makecell{Crafting a \\backdoored \\image \\encoder}}
	&	\multirow{3}{*}{\#epochs} 
		&	100 & 76.14 & 76.43 & 99.79  \\ \cline{3-6} 
	    && 150 & 76.14 & 76.45 & 99.84  \\ \cline{3-6} 
	    && 200 & 76.14 & 76.18 & 99.73  \\ \cline{2-6} 
	&	\multirow{3}{*}{Learning rate} 
	    & $1\times 10^{-3}$ & 76.14 & 76.18 & 99.73  \\ \cline{3-6} 
	    &&	$5\times 10^{-4}$ & 76.14 & 76.54 & 99.80 \\ \cline{3-6} 
	    && $1\times 10^{-4}$ & 76.14 & 75.53 & 97.76 \\ \cline{2-6}   
	&	\multirow{2}{*}{ Trigger} 
	    & A white square & 76.14 & 76.18 & 99.73  \\ \cline{3-6}  
	    &&	A random trigger & 76.14 & 75.83 & 100.0  \\ \hline
	    
	\multirow{9}{*}{\makecell{Training a \\ downstream \\classifier}}
	&	\multirow{3}{*}{ \#epochs} 
	    & 100 & 77.39 & 77.55 & 99.96 \\ \cline{3-6} 
	    && 300 & 76.14 & 76.39 & 99.80  \\ \cline{3-6} 
	    && 500 & 76.14 & 76.18 & 99.73  \\ \cline{2-6} 
	&	\multirow{3}{*}{Learning rate} 
	    & $1 \times 10^{-3}$ & 76.96 & 76.66 & 93.96 \\ \cline{3-6} 
	    && $5 \times 10^{-4}$ & 76.73 & 76.60 & 99.74  \\ \cline{3-6}  
	    &&	$1 \times 10^{-4}$ & 76.14 & 76.18 & 99.73 \\ \cline{2-6} 
 
	&	\multirow{3}{*}{\makecell{\#neurons \\in the two \\hidden layers} }
	    & [128,64] & 76.51 & 76.98 & 99.90  \\ \cline{3-6}  
	    && [256,128] & 76.13 & 76.64 & 99.90  \\ \cline{3-6} 
	    &&	[512,256] & 76.14 & 76.18 & 99.73 \\ \hline
	\end{tabular}
	}
	\label{result_of_different_parameters}
\vspace{-5mm}
\end{table}

\begin{figure}[!htbp]
	 \centering
\subfloat[CIFAR10]{\includegraphics[width=0.12\textwidth]{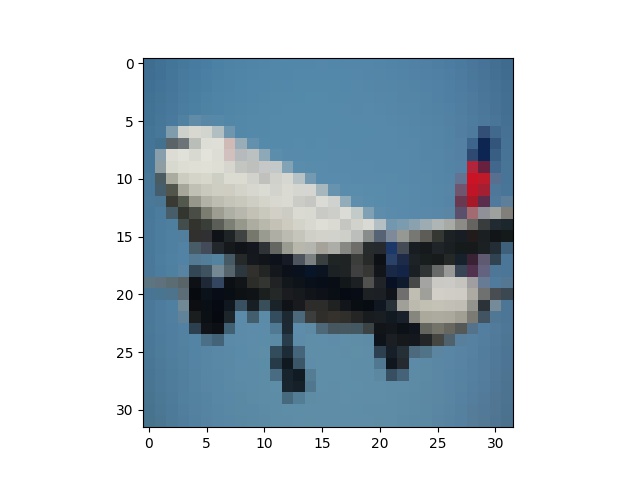}}
\subfloat[SVHN]{\includegraphics[width=0.12\textwidth]{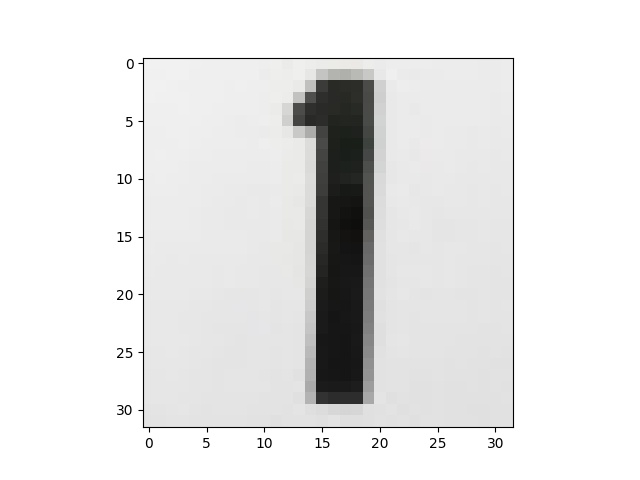}}
\subfloat[GTSRB]{\includegraphics[width=0.12\textwidth]{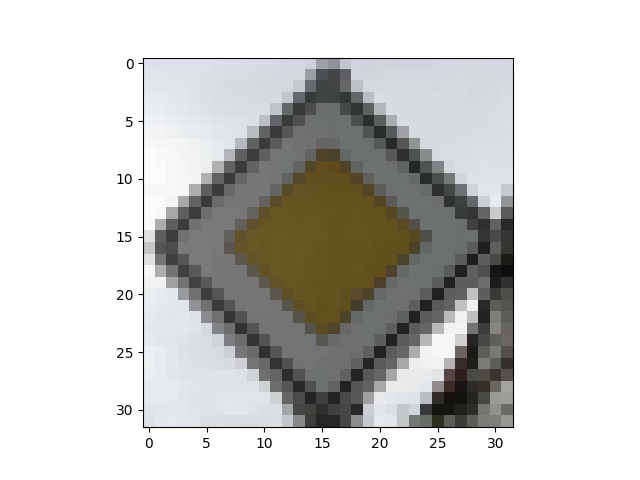}}
\subfloat[STL10]{\includegraphics[width=0.12\textwidth]{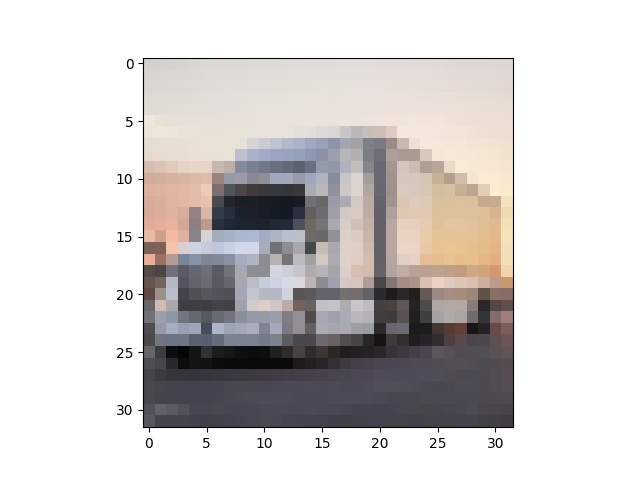}}
\caption{The default reference inputs for CIFAR10, SVHN, GTSRB, and STL10.}
\label{four_attack_inputs}
\vspace{-5mm}
\end{figure}

\begin{figure}[!htbp]
	 \centering
\subfloat[Truck 0]{\includegraphics[width=0.15\textwidth]{figs/truck.jpg}}
\subfloat[Truck 1]{\includegraphics[width=0.15\textwidth]{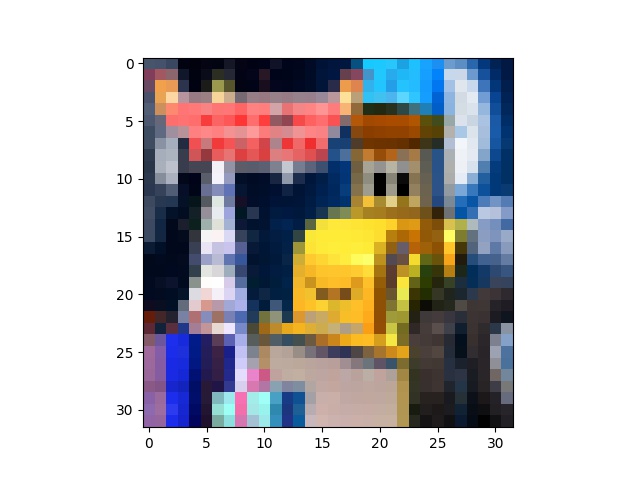}}
\subfloat[Truck 2]{\includegraphics[width=0.15\textwidth]{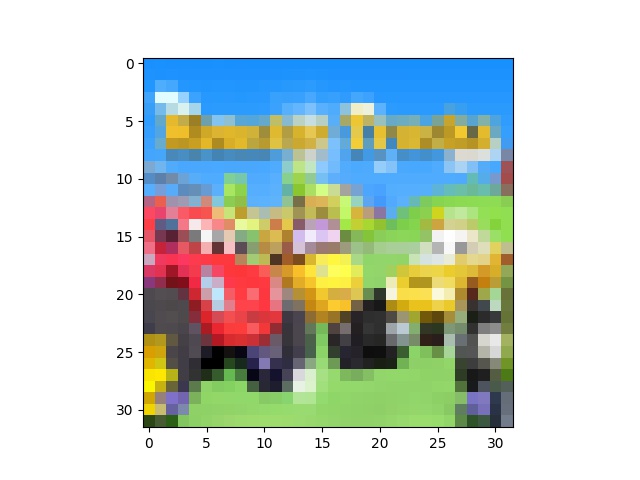}}
\caption{The reference inputs for STL10 used in Table~\ref{multiple_attack_inputs}. Truck 0 is the default reference input.}
\label{multiple_classes_attack_inputs}
\end{figure}

\begin{figure}[!htbp]
	 \centering
\subfloat[GTSRB]{\includegraphics[width=0.24\textwidth]{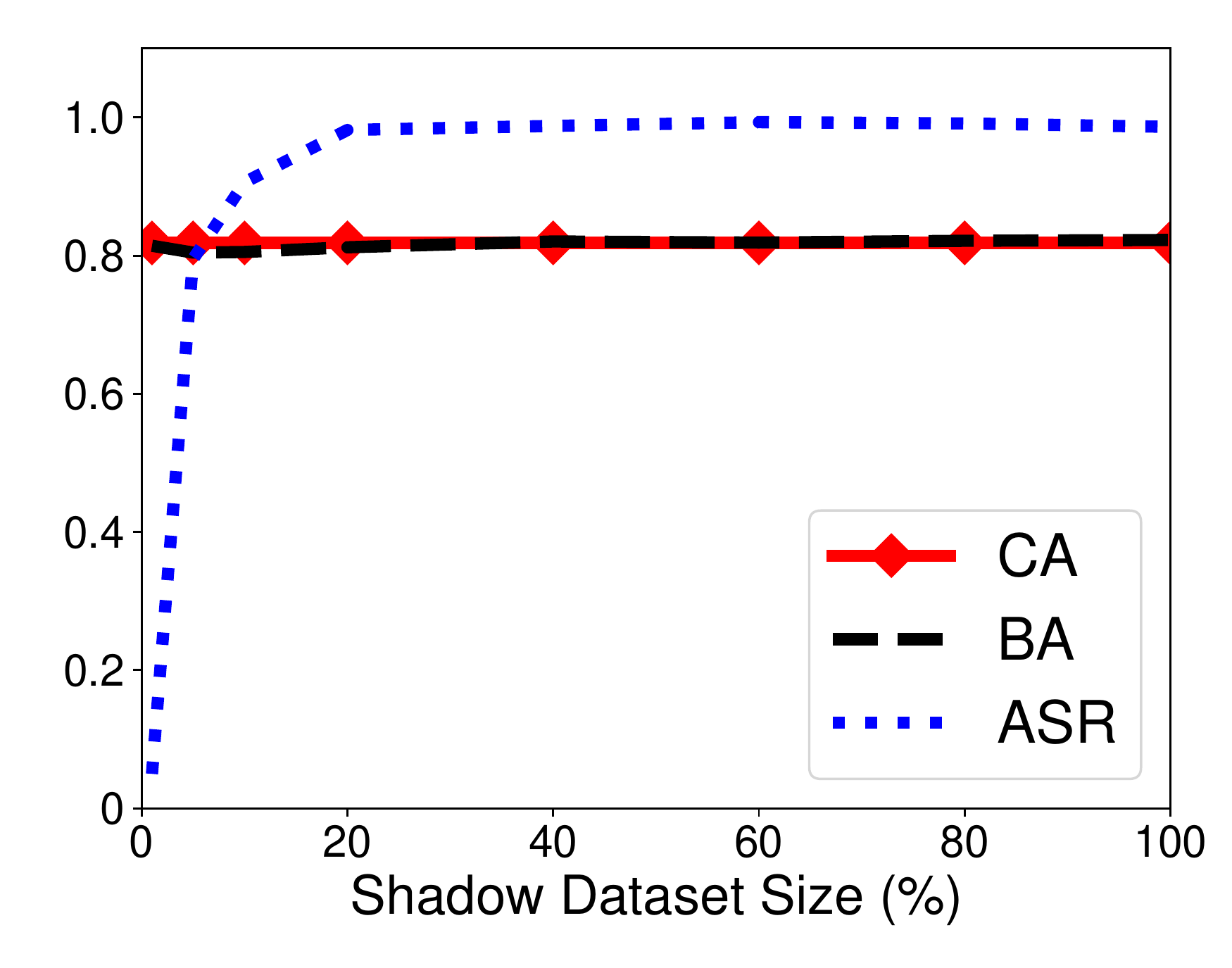}}
{\includegraphics[width=0.24\textwidth]{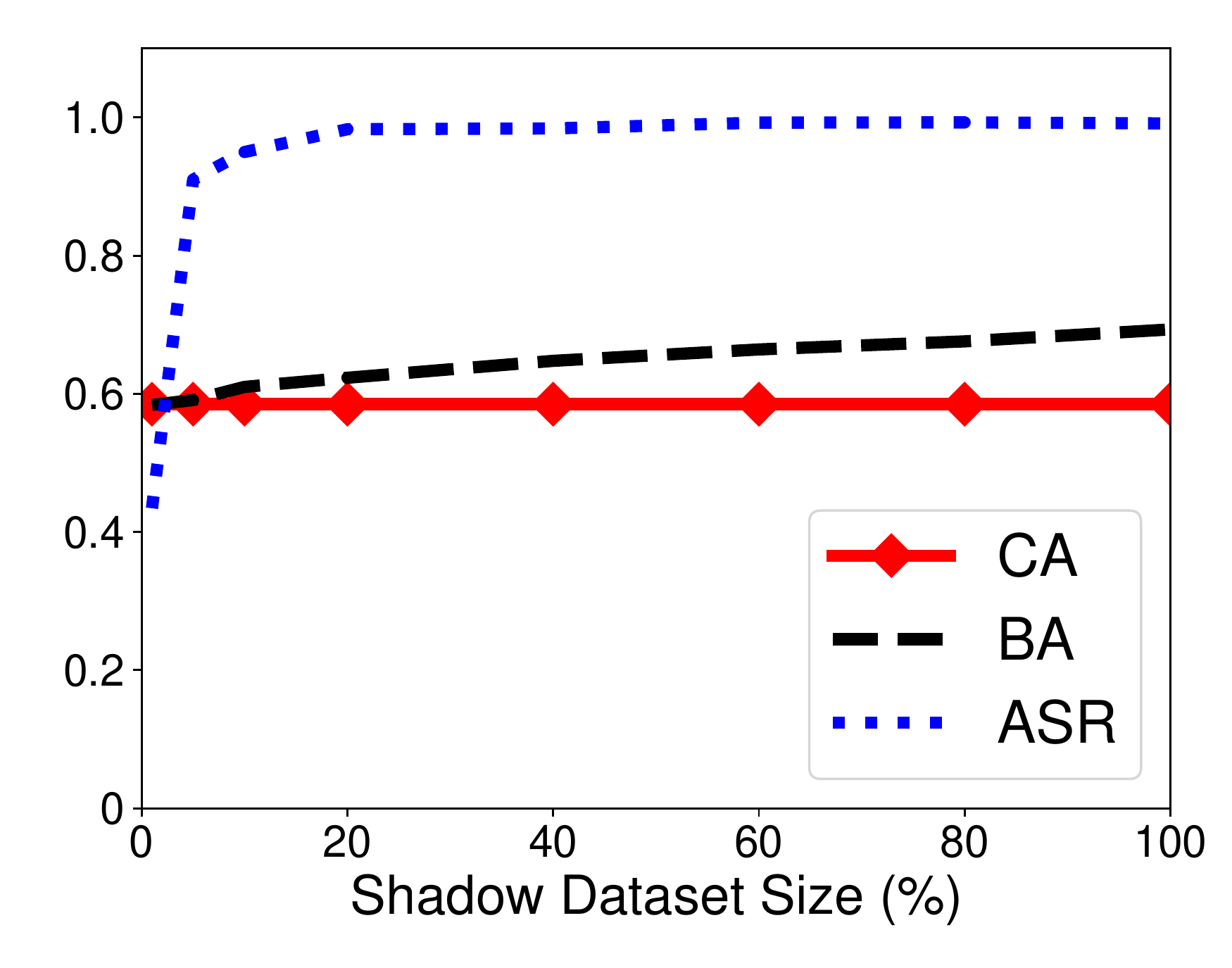}}
\caption{The impact of the shadow dataset size on our BadEncoder when the target downstream datasets  are GTSRB  (left) and SVHN (right). The shadow dataset is a subset of the pre-training dataset, which is CIFAR10.}
\label{impact_of_attack_dataset_size_svhn_stl10}
\vspace{-3mm}
\end{figure}

\begin{figure*}[!htbp]
	 \centering
\subfloat[GTSRB]{\includegraphics[width=0.32\textwidth]{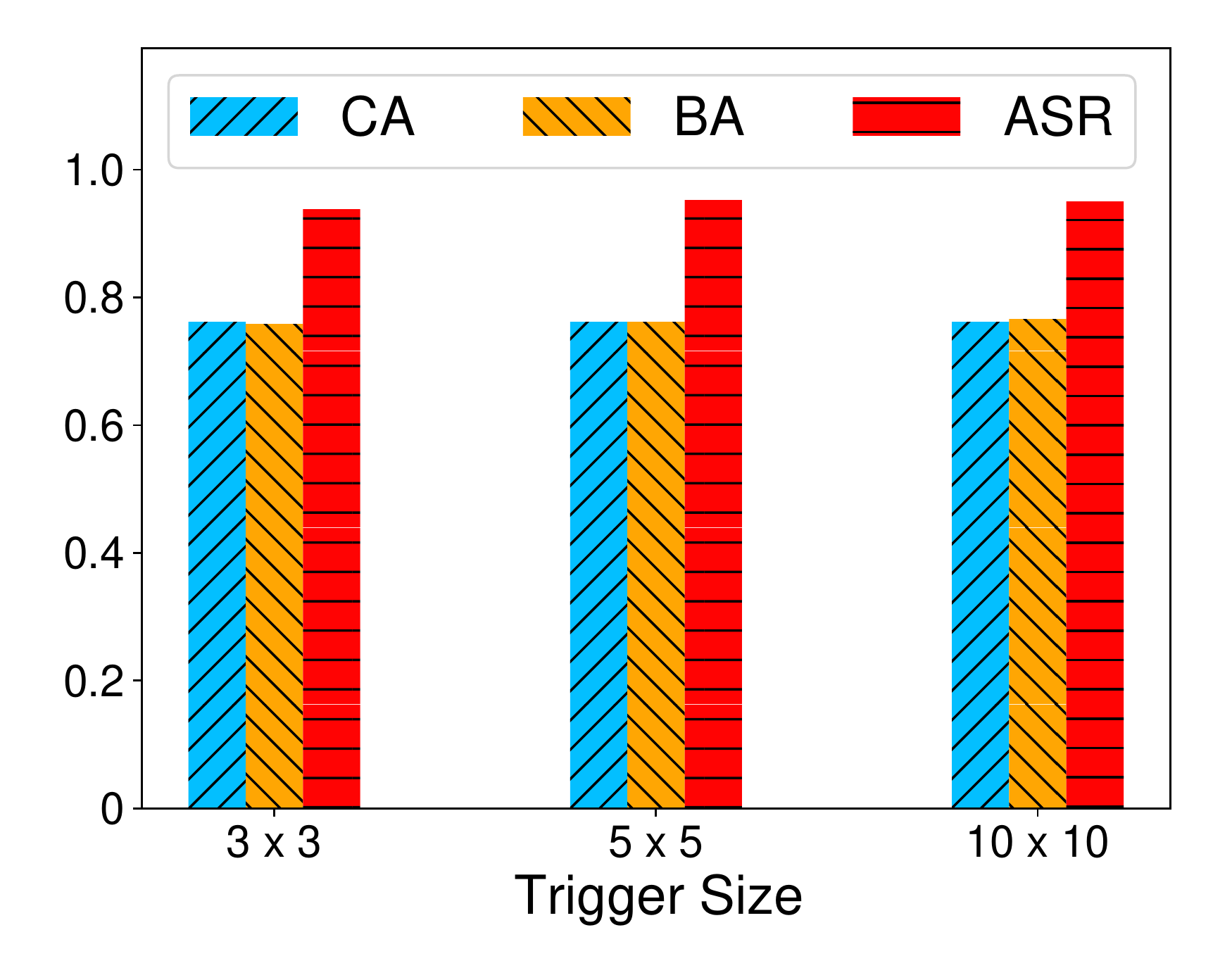}}
\subfloat[SVHN]{\includegraphics[width=0.32\textwidth]{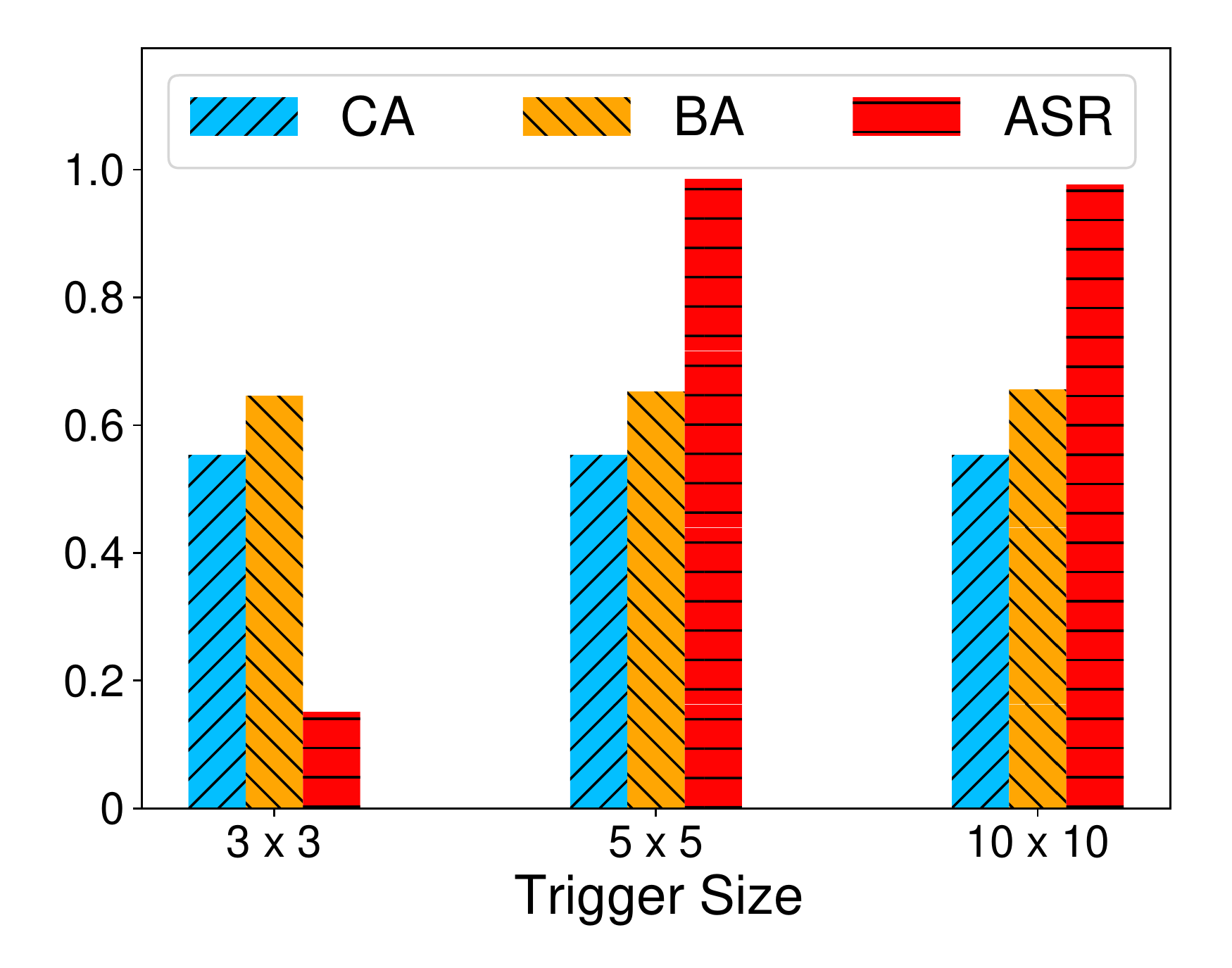}}
\subfloat[CIFAR10]{\includegraphics[width=0.32\textwidth]{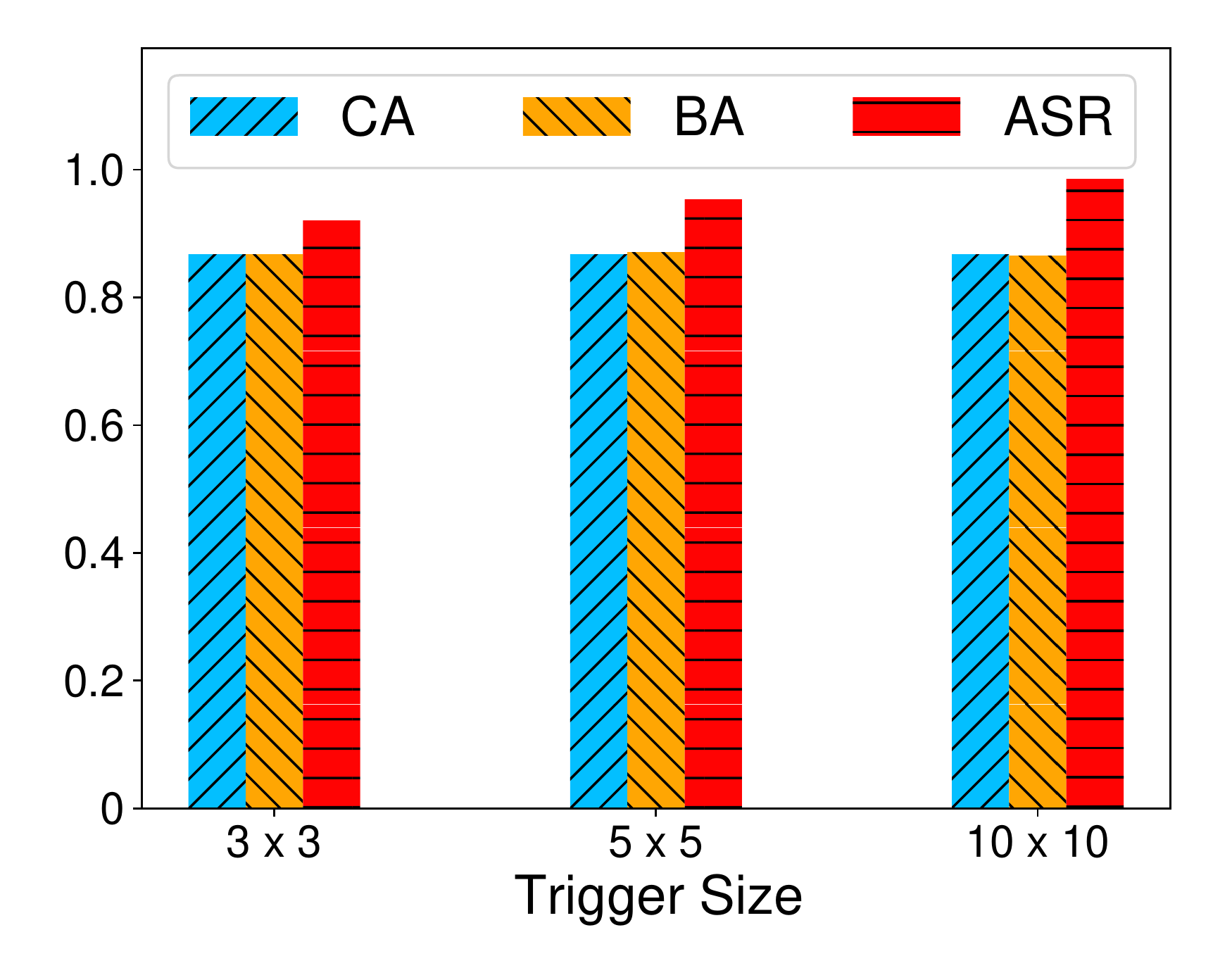}}
\caption{The impact of the trigger size on our BadEncoder for different target downstream datasets when the pre-training dataset is STL10.}
\label{impact_of_trigger_size_stl10}
\end{figure*}

\begin{figure}[!htbp]
	 \centering
\subfloat[Airplane]{\includegraphics[width=0.15\textwidth]{figs/airplane_orig.jpg}}
\subfloat[Truck]{\includegraphics[width=0.15\textwidth]{figs/truck.jpg}}
\subfloat[Horse]{\includegraphics[width=0.15\textwidth]{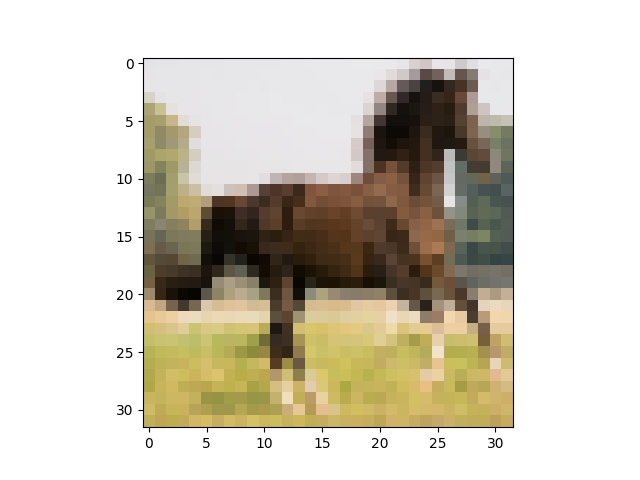}}
\caption{The reference inputs for attacking three target classes in STL10 simultaneously, which are used in Table~\ref{multiple_target_classes}.}
\label{multiple_downstream_inputs}
\end{figure}

\begin{figure}[!htbp]
	 \centering
\subfloat[SVHN]{\includegraphics[width=0.15\textwidth]{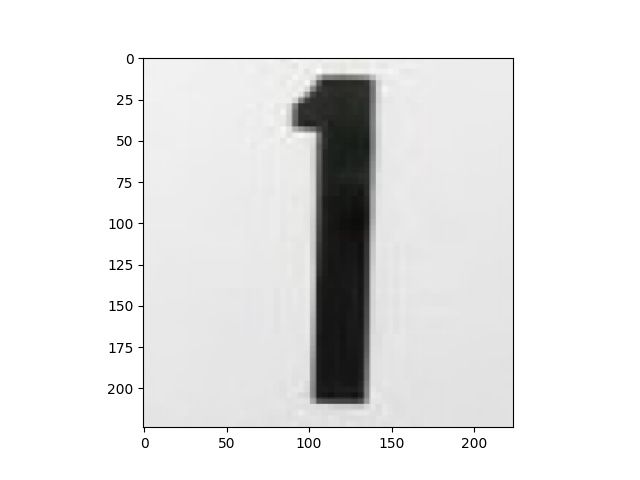}}
\subfloat[GTSRB]{\includegraphics[width=0.15\textwidth]{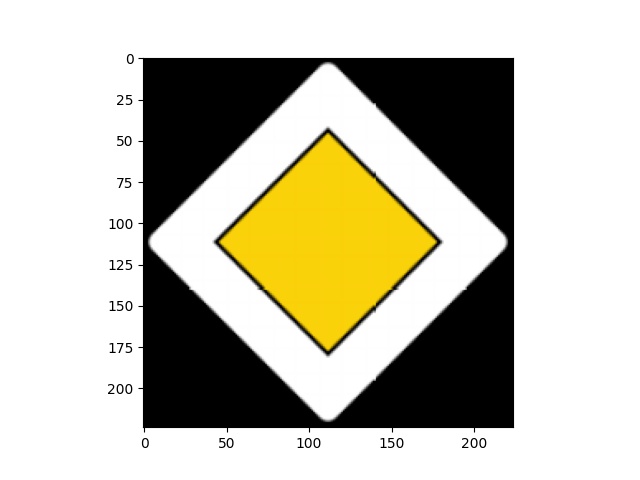}}
\subfloat[STL10]{\includegraphics[width=0.15\textwidth]{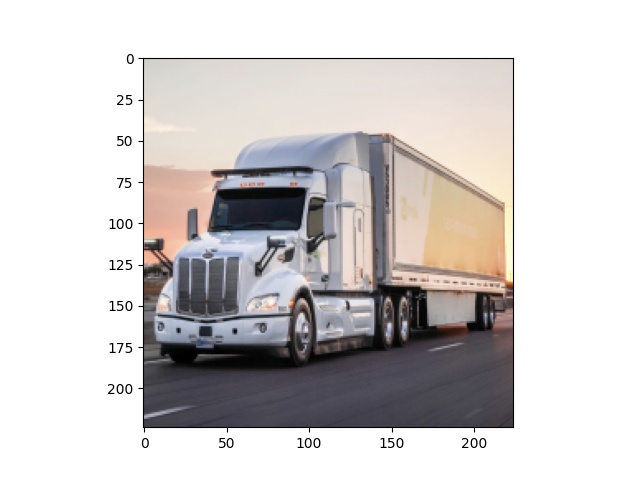}}
\caption{The reference inputs for attacking the image encoder pre-trained on ImageNet  by Google, which are used in Table~\ref{case_study_resnet50_google}.}
\label{resnet50_google_attack_input}
\end{figure}

\begin{figure}[!htbp]
	 \centering
\subfloat[SVHN]{\includegraphics[width=0.15\textwidth]{figs/one_1_224.jpg}}
\subfloat[GTSRB]{\includegraphics[width=0.15\textwidth]{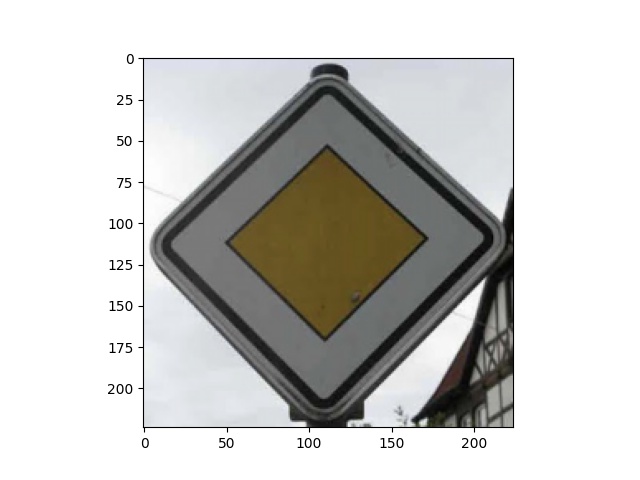}}
\subfloat[STL10]{\includegraphics[width=0.15\textwidth]{figs/truck_224.jpg}}
\caption{The reference inputs for attacking  OpenAI's CLIP  in the multi-shot classifier scenario, which are used in Table~\ref{case_study_clip_st_1}.}
\label{resnet50_clip_attack_input}
\end{figure}

\begin{figure}[!t]
	 \centering
\subfloat[SVHN]{\includegraphics[width=0.15\textwidth]{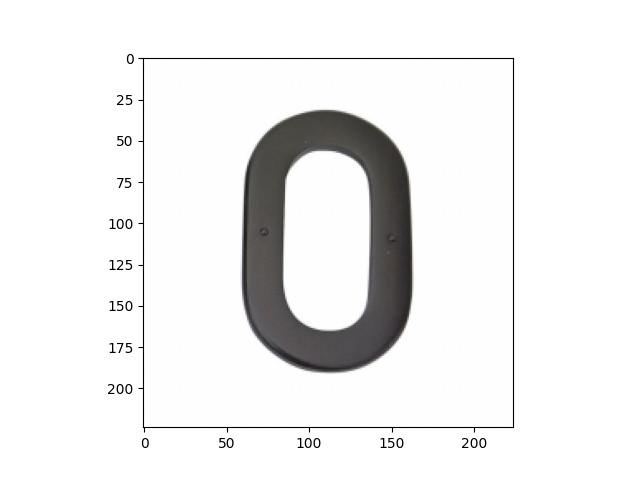}}
\subfloat[GTSRB]{\includegraphics[width=0.15\textwidth]{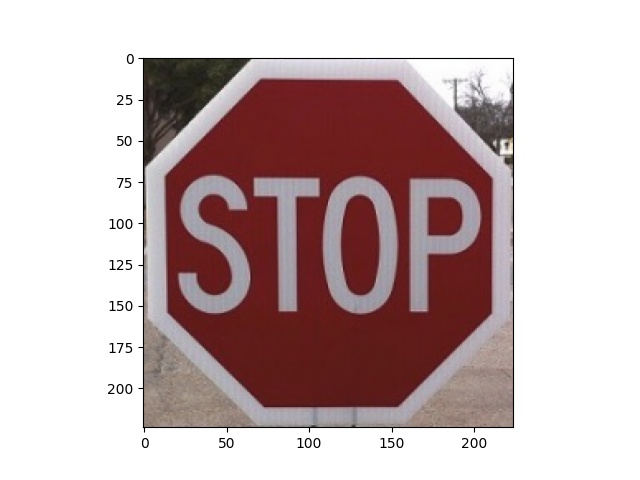}}
\subfloat[STL10]{\includegraphics[width=0.15\textwidth]{figs/truck_224.jpg}}
\caption{The reference inputs for attacking  OpenAI's CLIP in the zero-shot classifier scenario, which are used in Table~\ref{case_study_clip_zs}.}
\label{clip_zero_shot_attack_input}
\end{figure}

\begin{figure}[!t]
	 \centering
\subfloat[SVHN]{\includegraphics[width=0.24\textwidth]{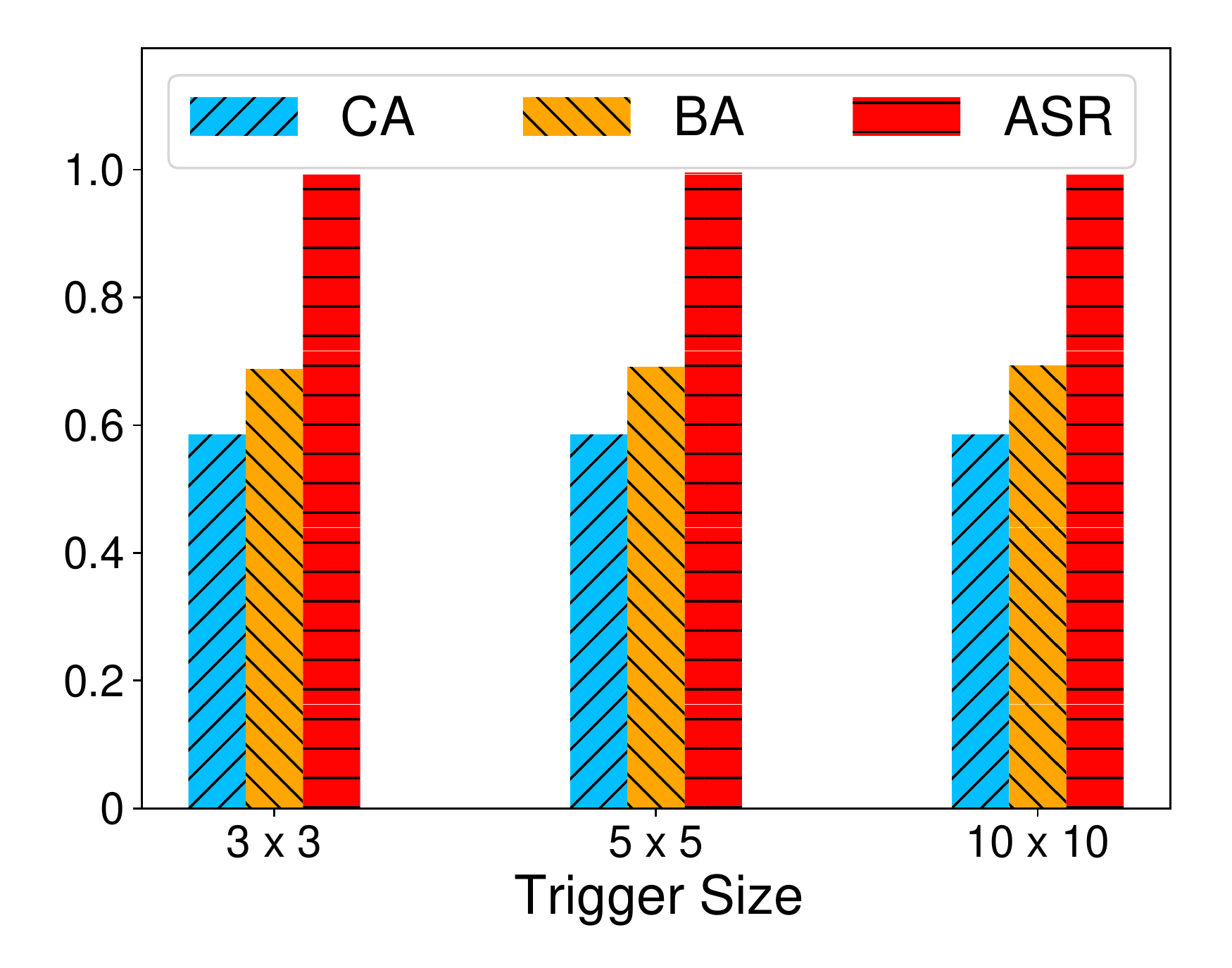}}
\subfloat[STL10]{\includegraphics[width=0.24\textwidth]{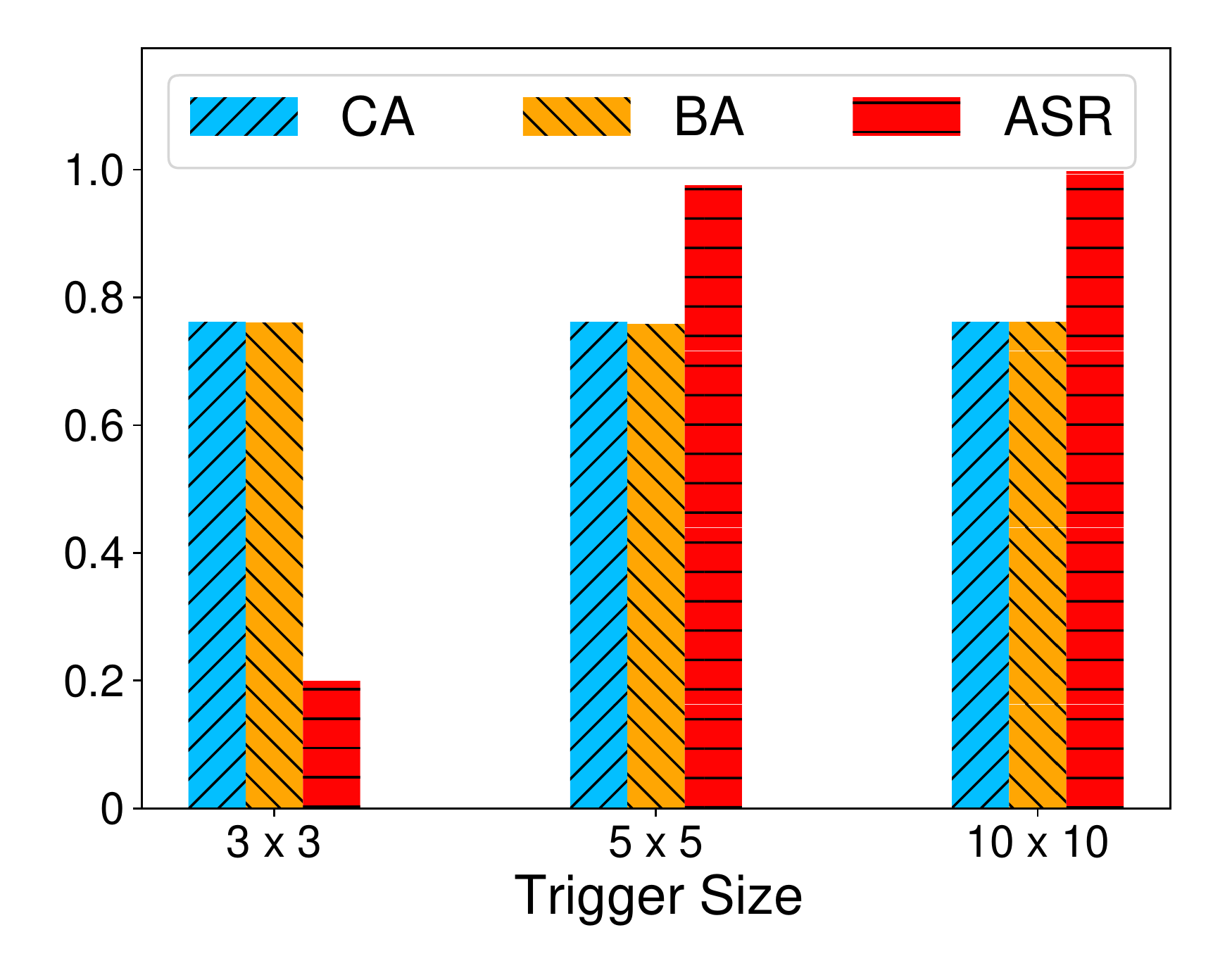}} 
\caption{The impact of the trigger size for SVHN and STL10. The pre-training dataset is CIFAR10.}
\label{impact_of_trigger_size_cifar10}
\vspace{-2mm}
\end{figure}

\end{document}